\pdfoutput=1
\documentclass[aps,preprint,onecolumn,superscriptaddress]{revtex4-1}

\usepackage{graphicx} % Allows for eps images
\usepackage{epsfig}
\usepackage{amsmath} % AMS Math Package
\usepackage{amsthm} % Theorem Formatting
\usepackage{amssymb}	% Math symbols such as \mathbb
\usepackage{physics}
\usepackage{graphics} % or graphicx 
\usepackage{hyperref} %clickable references.
\hypersetup{
    colorlinks,
    citecolor=blue,
    filecolor=black,
    linkcolor=red,
    urlcolor=blue
}

\usepackage{bm}
\usepackage{epstopdf}
\usepackage{array}
\usepackage{enumitem}
\usepackage{verbatim} 
\usepackage{epsfig}
\usepackage{tikz}
\usepackage{float}
\usepackage{makecell}
\usepackage{booktabs}
\usepackage{setspace}

\begin{document}

\title{The limits of multifunctionality in tunable networks}
\author{Jason W. Rocks}
\thanks{J.W.R. and H.R. contributed equally to this work.}
\affiliation{Department of Physics and Astronomy, University of Pennsylvania, Philadelphia, PA 19104, USA}
\author{Henrik Ronellenfitsch}
\thanks{J.W.R. and H.R. contributed equally to this work.}
\affiliation{Department of Mathematics, Massachusetts Institute of Technology, Cambridge, MA 02139, USA}
\author{Andrea J. Liu}
\affiliation{Department of Physics and Astronomy, University of Pennsylvania, Philadelphia, PA 19104, USA}
\author{Sidney R. Nagel}
\affiliation{The James Franck and Enrico Fermi Institutes and The Department of Physics,
The University of Chicago, Chicago, IL 60637, USA}
\author{Eleni Katifori}
\affiliation{Department of Physics and Astronomy, University of Pennsylvania, Philadelphia, PA 19104, USA}

\begin{abstract}
Nature is rife with networks that are functionally optimized to propagate inputs in order to perform specific tasks. Whether via genetic evolution or dynamic adaptation, many networks create functionality by locally tuning interactions between nodes. Here we explore this behavior in two contexts: strain propagation in mechanical networks and pressure redistribution in flow networks. By adding and removing links, we are able to optimize both types of networks to perform specific functions. We define a single function as a tuned response of a single ``target" link when another, predetermined part of the network is activated. Using network structures generated via such optimization, we investigate how many simultaneous functions such networks can be programmed to fulfill. We find that both flow and mechanical networks display qualitatively similar phase transitions in the number of targets that can be tuned, along with the same robust finite-size scaling behavior. We discuss how these properties can be understood in the context of a new class of constraint-satisfaction problems.
\end{abstract}
\maketitle

\section*{Introduction}
Many naturally occurring and synthetic networks are endowed with a specific and efficient
functionality.
For example, allosteric proteins globally adjust their conformation upon binding a ligand in order to
control the activity of a distant active site~\cite{Motlagh2014,Ribeiro2016}.
Gene regulatory networks express specific proteins~\cite{Barabasi2004}.  Biological and artificial neural networks retrieve memories based on a limited number of inputs~\cite{Hertz1991,McCulloch1943,Hopfield1982}.
In some cases, networks can adapt and change their function depending on the needs of the system; venation networks in plants~\cite{Katifori2010,Pitterman2010}, animals~\cite{Hu2013,Ronellenfitsch2016,LaBarbera1990}, and slime molds~\cite{Tero2010,Alim2013} can reroute the transport of fluids, enhancing or depleting nutrient levels in order to support local growth or activity. Modern power grids must precisely distribute electrical energy generated from a limited number of sources to a large number of consumers with widely varying consumption needs at different times~\cite{Fang2012}. All of these networks are optimized to some degree, either by evolution via natural selection, dynamic reconfiguration, or by human planning and ingenuity.

A key aspect of such functionality is the complexity of a specific task.  We define a ``function'' as an optimized response of a localized component of a network when another predefined, localized component of the system is activated. A ``task'' is then defined as the collective response of a set of individual functions due to a single input. The number of functions representing a specific task is the task complexity.

In this work we address the limits of complexity for a single task: that is, how many functions comprising a single task can be programmed into a network? We consider two examples: (i) mechanical networks - in which nodes are connected by central-force harmonic springs - locally flexing in response to an applied strain and (ii) flow (or resistor) networks - in which nodes are connected by linear resistors - locally producing a pressure drop due to an applied pressure at the source. These systems are related; flow networks are mathematically equivalent to mechanical networks embedded in one spatial dimension - but with a nontrivial node topology~\cite{Tang1987}.

%The random removal of springs in a mechanical network has little effect on many macroscopic properties, such as the ratio of the shear and bulk moduli~\cite{Ellenbroek2009}. In contrast, targeted removal of only a select tiny fraction of the springs can tune this ratio to any desired value or create other macroscopic functionalities in the network~\cite{Goodrich2015, Hexner2018, Reid2018}.
The macroscopic properties of mechanical networks, such as their bulk and and shear moduli, can be finely tuned by modifying only a select tiny fraction of the springs between nodes~\cite{Goodrich2015, Hexner2018, Reid2018} (in contrast to random removal~\cite{Ellenbroek2009}). 
Previously, this idea was extended to show that such networks can be tuned to develop allosteric behavior via selective spring removal~\cite{Rocks2017,Yan2017, Flechsig2017}.
Allostery in these systems corresponds to designing a task composed of a single function in which a randomly selected spring (the target) responds in a specified way to a strain imposed on a separate pair of nodes (the source). Here we further develop this idea by simultaneously tuning multiple targets controlled by a single source in both elastic and flow networks.
We investigate the question of how many individual targets can be tuned successfully (i.e., the scaling of the maximal task complexity) as a function of the size of the network.
We find that in both flow and mechanical networks, the limit of task complexity is set by a phase transition.

\section*{Network Tuning Protocol}

\begin{figure}[h!]
\centering
\includegraphics[width=0.75\linewidth]{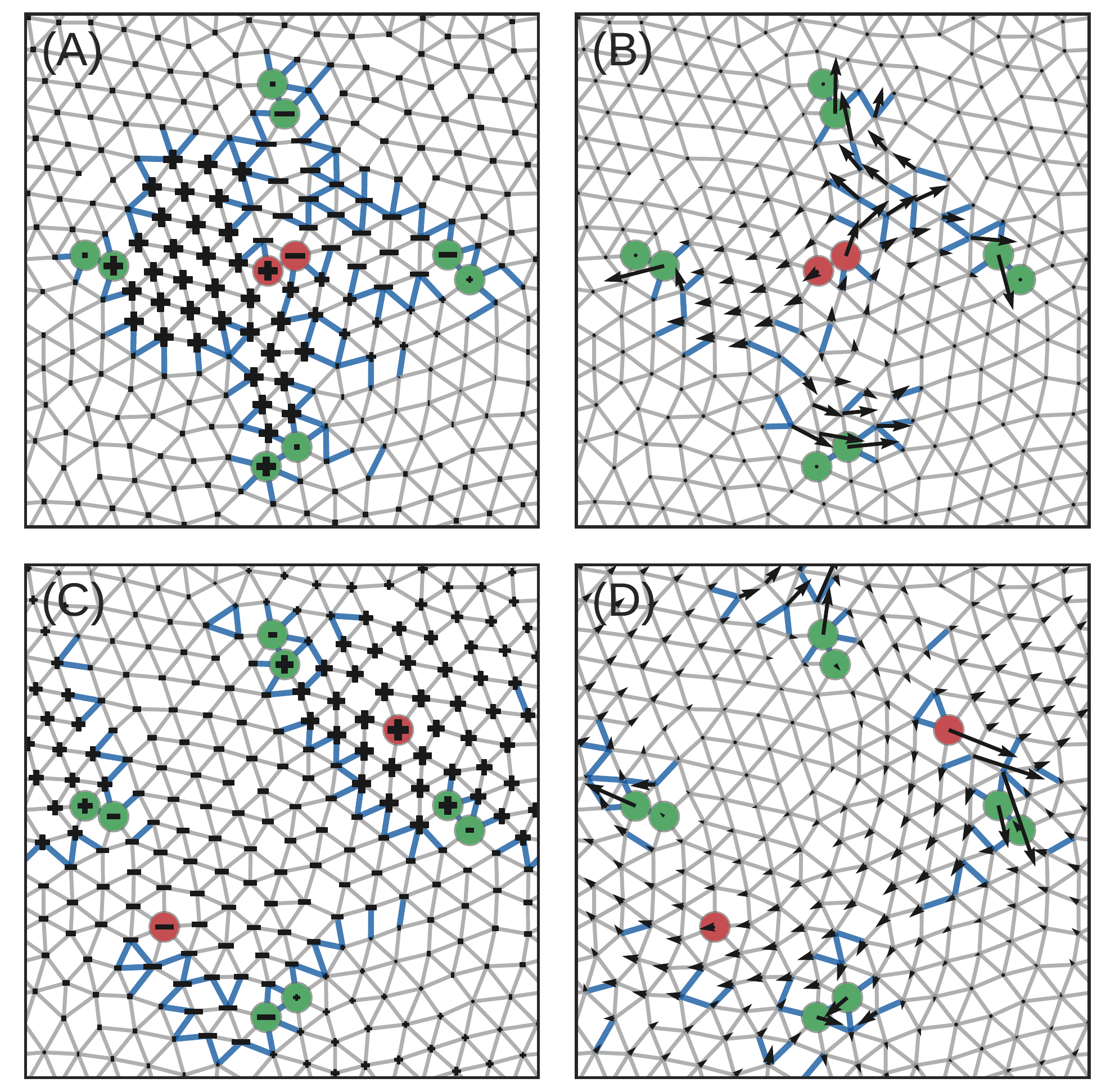}
\caption{Networks tuned to display multifunctional responses. Each network starts with the same initial topology and same choice of four target edges (corresponding nodes shown in green).  In (A) and (C) the network carries flow, while in (B) and (D) it is a two-dimensional mechanical network. For each network a source extension (pressure drop) is applied to a pair of source nodes (shown in red). In (A) and (B) the  pair of source nodes is connected by an edge, while in (C) and (D) the source  nodes are not connected by an edge. For flow networks, response ratios have been tuned to $\eta_\alpha \geq 0.5$, while for the mechanical networks they are $\eta_\alpha \geq 1.0$. The edges removed by the tuning algorithm are shown as thick blue lines. For flow networks, the resulting pressure magnitude of the tuned network is indicated by the size of the black nodes, while the sign of the pressure is represented by the node shape. For mechanical networks, the resulting node displacements are shown as black arrows. }

\label{fig:networks}
\end{figure}

Our method for tuning networks follows the general scheme described in our previous work~\cite{Rocks2017} with some slight modifications.
We start with two-dimensional configurations of soft spheres with periodic boundary conditions created using standard jamming algorithms.
We extract the particle contact structure by placing nodes at the centers of each sphere and links (edges) between nodes corresponding to overlapping particles. This ensemble of networks is used for both spring and flow networks.
For spring networks, edges are unstretched central-force springs, while for flow networks, edges are resistive conduits. By using the same set of nodes and edges for both systems, we can directly compare results.

For each network, a pair of source nodes is chosen randomly, along with a set of $N_T$ target edges. Our goal is to tune the extension (or pressure drop) $e_\alpha$ of each target edge, labeled by $\alpha$, in response to an extension (pressure drop) $e_S$ applied to the source nodes by adding and removing edges from the network. We explore two different types of sources: pairs of nodes connected by a randomly chosen edge and pairs of nodes that are each chosen randomly anywhere in the network (see SI Appendix for global compression and shear sources in mechanical networks).

To control the response of the targets, we define the response ratio $\eta_\alpha \equiv e_\alpha / e_S$ for each target. Each $\eta_\alpha$ is in general a collective property of the network; the response of each target is a function of the total network structure. Before tuning the network, we measure the  initial extension (pressure drop) $e^{(0)}_\alpha$ to obtain the initial response ratio of each target $\eta^{(0)}_\alpha = e^{(0)}_\alpha / e_S$. We then tune the response ratio of each target so that its relative change as compared to the initial state is greater than or equal to a specified positive constant $\Delta$; that is, we tune each response ratio to satisfy the constraint
\begin{align}
\frac{\eta_\alpha - \eta^{(0)}_\alpha}{\eta^{(0)}_\alpha} \geq \Delta, \quad  \alpha = 1,\ldots,N_T .\label{eq:constraint}
\end{align}
Thus, for mechanical networks we require contracting edges to contract more, and
expanding edges to expand more. For flow networks, we require the magnitude of the pressure drop to increase without changing the direction of the flow through each target link.

Our optimization scheme involves minimizing a loss function which penalizes deviations from the constraints in $\eqref{eq:constraint}$ (see Methods and Materials). Each optimization step consists of either removing a single link, or reinserting a previously removed link to modify the network topology in discrete steps. More specifically, at each step we measure the resulting change in the loss function for each possible single link removal or reinsertion and then remove or reinsert the link which minimizes the loss function at that step.

Fig.~\ref{fig:networks} depicts examples of both flow and mechanical networks which have been tuned using our prescribed method for the two different types of applied sources. Fig.~\ref{fig:networks}(A) and (B) show flow and mechanical networks, respectively, tuned to respond to a source applied to a pair of nodes connected by an edge. Fig.~\ref{fig:networks}(C) and (D) show the same networks, but with a pair of source nodes that are not connected by an edge.

\section*{Results}

\begin{figure*}[h!]
\centering
\includegraphics[width=\linewidth]{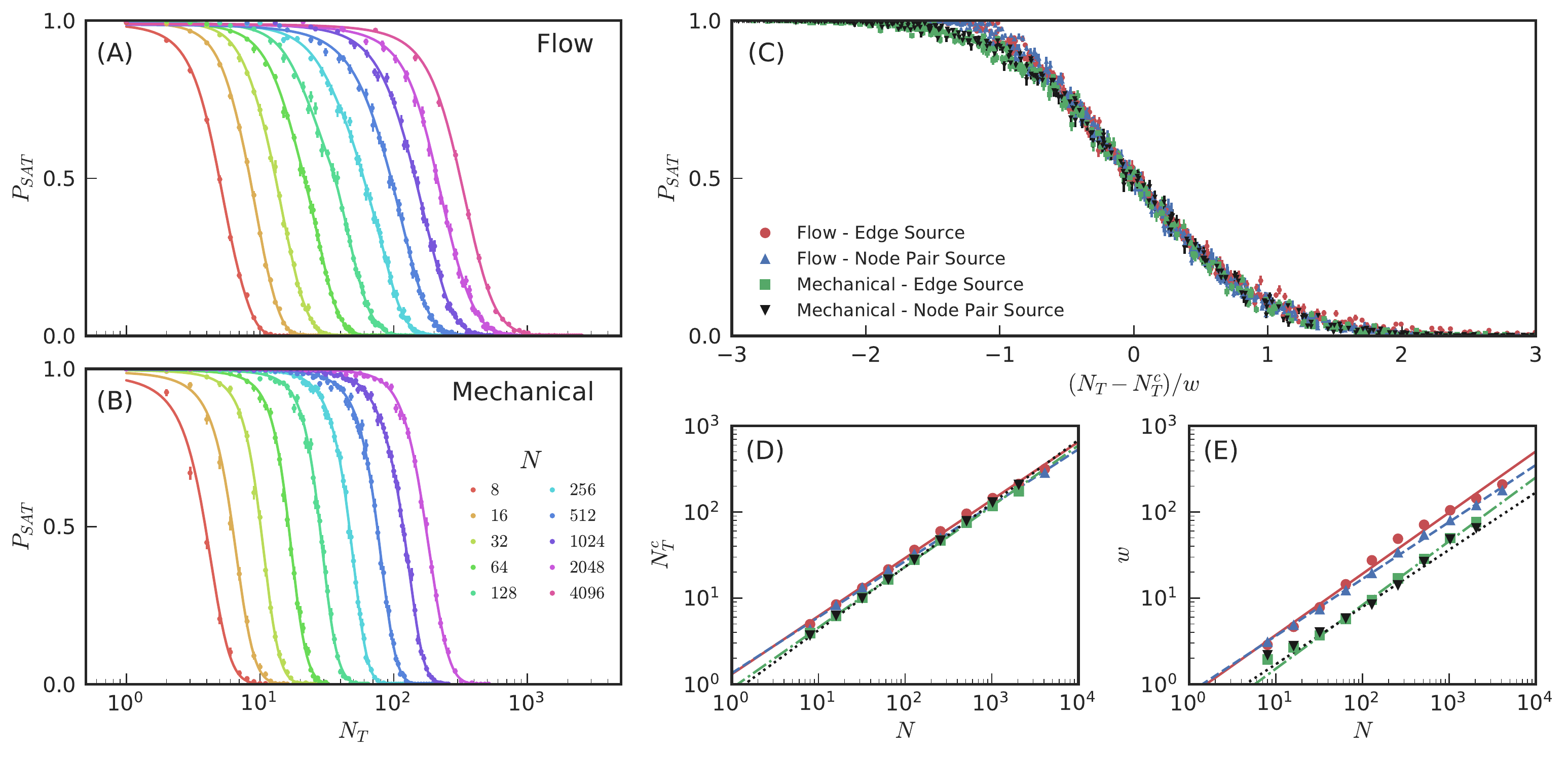}
\caption{The fraction of satisfied configurations $P_{SAT}$ for (A) flow networks and (B) two-dimensional mechanical networks as a function of number of targets $N_T$ for systems of $N$ nodes. Results are shown for a pressure or extension applied to a single source edge with a desired relative change in target response of $\Delta = 0.1$. Smoothing splines are shown as an estimate of the underlying satisfiability transition, while estimated error bars are shown for binomially distributed data (see SI Appendix). 
(C) Scaling collapse of all systems sizes for four different cases: flow networks with an edge source (red circles) and with a node pair source (blue triangles) and similarly, mechanical networks with an edge source (green squares) and with a node pair source (black triangles). In each case, we plot $P_{SAT}$ vs. $(N_T-N_T^c)/w$ where $P_{SAT} = \frac{1}{2}$ at $N_T^c$ and $w$ is the interval in $N_T$ over which $0.25 < P_{SAT} < 0.75$.  (D) The transition points $N_T^C$ and (E) width of the transition $w$ are reasonably described by power laws as a function of $N$. The power law fits for $N_T^c$ have exponents of approximately $0.67$ and $0.65$ for flow networks and $0.71$ and $0.74$ for mechanical networks with an edge and node pair source, respectively. In the same order, the power law fits for $w$ have exponents of $0.71$, $0.66$, $0.74$, and $0.66$.}
\label{fig:psat}
\end{figure*}

\begin{figure}[h!]
\centering
\includegraphics[width=0.75\linewidth]{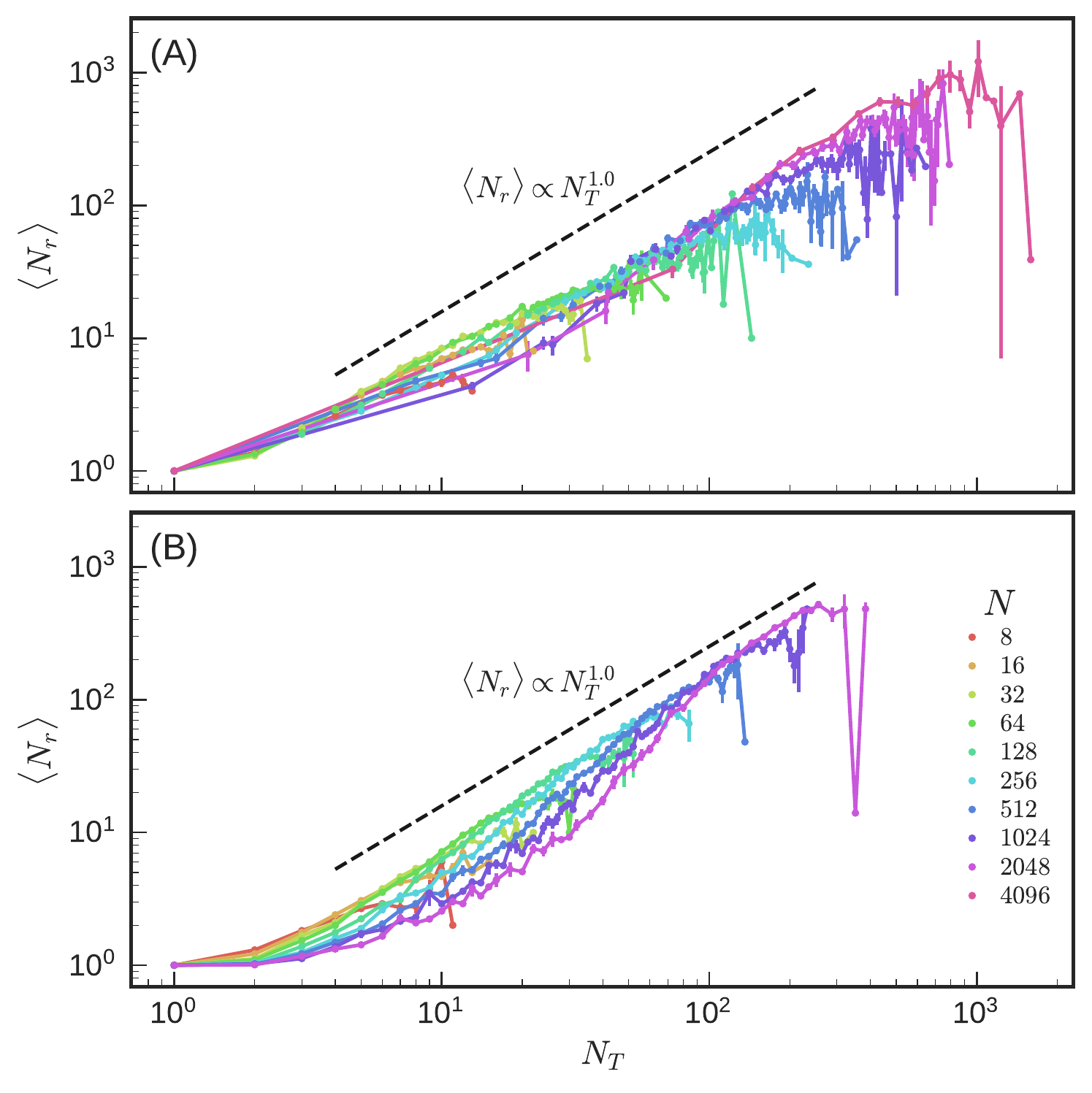} \\
\caption{Power law behavior of the average number of removed edges as a function of number of targets $N_T$ for (A) flow networks and (B) mechanical networks for various system sizes $N$. Included networks correspond to those that have been tuned successfully in Figs.~\ref{fig:psat}(A) and (B) with an edge source and desired change in target response of $\Delta = 0.1$. Error bars indicate the error on the mean. Power laws with an exponent of $1.0$ are depicted as black dashed lines for comparison.}
\label{fig:removals}
\end{figure}

We investigate the ease with which networks can be tuned as a function of the number of targets, i.e. the task complexity.
For both flow and mechanical networks, we explore the effects of various aspects of the tuning problem.
Figs.~\ref{fig:psat}(A) and (B) display typical results for the fraction of networks that can be tuned successfully, $P_{SAT}$,
for flow and mechanical networks, respectively.
Data is shown for a randomly chosen edge source and $N_T$ randomly chosen target edges with a desired relative change in target response of $\Delta = 0.1$.
System sizes range from $N=8$ to $4096$ nodes. Each value of $P_{SAT}$ is calculated by tuning at least 512 independent randomly generated networks. At low $N_T$, $P_{SAT} \approx 1$ while at large $N_T$, $P_{SAT}$ drops to zero. 

In Fig.~\ref{fig:psat}(C), we plot the transition curves for all system sizes for the four cases studied on the same axes. Using the smoothing spline interpolations shown in Fig.~\ref{fig:psat}(A) and (B) (see SI Appendix), we estimate the number of targets $N_T^c$ at which $P_{SAT} = 0.5$. Next, we estimate the width of the transition, $w$, taken as the interval in $N_T$ over which $0.25 < P_{SAT} < 0.75$. We attempt to collapse each curve by plotting $P_{SAT}$ vs. $(N_T-N_T^c)/w$. We find a similar functional form for all cases, with only a slight difference between flow and mechanical networks near $(N_T-N_T^c)/w \approx -1$.

Figs.~\ref{fig:psat}(D) and (E) show that flow networks and mechanical networks have similar power-law behaviors for $N_T^c$ and $w$.  Both the transition location and width scale approximately as $N^\nu$ with $\nu \approx 0.7$.  Because the scaling exponent for $N_T^c$ is less than 1, the critical \emph{fraction} of functions that can be tuned simultaneously approaches zero as $N$ goes to infinity, even though the \emph{number} of simultaneously tuned functions diverges with system size.
Thus small networks are relatively more tunable than large ones.
In addition, the sub-linear scaling of the transition width shows that $P_{SAT}$ drops more rapidly with $N_T/N$ as $N$ increases, implying that the crossover becomes sharp as $N \rightarrow \infty$. At the same time, Fig.~\ref{fig:removals} shows that the average number of links that need to be removed for a successful tuning operation grows approximately linearly with the number of targets. Thus, those networks that can be tuned successfully typically only require removal of a constant fraction of edges. Together, our results suggest $P_{SAT}(N_T/N) \sim F[(N_T/N - \rho^\infty) N^{1-\nu}]$  with $\rho^\infty$ consistent with zero. For $\nu < 1$, this implies a random first order transition in the thermodynamic limit, with a discontinuity in $P_{SAT}$ and power-law finite-size scaling. Such hybrid transitions are typical of constraint-satisfaction problems.

\section*{Discussion}

We framed the problem of the maximum number of target edges that can be tuned successfully in a mechanical or flow network as a type of discrete constraint-satisfaction problem, in which we asked how many inequality constraints can be satisfied simultaneously. This places the tuning of multifunctionality in the context of a variety of other problems in physics, mathematics, and computer science,
including jamming~\cite{Ohern2003}, spin glasses~\cite{Berthier2011},
the $k$-SAT problem~\cite{Mezard2002}, $k$-core percolation~\cite{Schwarz2006}, and the perceptron~\cite{Franz2017}.
Much progress has been made by linking such transitions to the statistical physics of critical phenomena. The hallmark of these systems is the emergence of a SAT-UNSAT transition between regions in parameter
space where the constraints can always (or with high probability) be satisfied
and regions where the system is \emph{frustrated}, such that not all constraints can be satisfied
simultaneously~\cite{Franz2017}.
In mean-field, and in some cases in finite dimensions, the SAT-UNSAT transition is a random first-order transition, with a discontinuous jump in the order
parameter (the fraction of satisfied configurations $P_{SAT}$) as in a first-order phase transition, but with power law scaling as in a second-order transition.

We have demonstrated a SAT-UNSAT transition in the complexity of a single task that can be tuned into disordered mechanical and flow networks. In both cases, the maximum task complexity diverges with a power law that is sublinear in $N$, the number of nodes in the network. The width of the SAT-UNSAT transition (relative to $N$) vanishes as $N$ diverges, showing that the transition is a true phase transition.

Although we find $P_{SAT}(N_T/N) \sim F[(N_T/N - \rho^\infty) N^{1-\nu}]$ for the four cases displayed in Fig.~\ref{fig:psat}, both $F(x)$ and $\nu$ can vary depending on a variety of factors. These factors include: (i) the local or global nature of the source, (ii) the magnitude of desired change in target response $\Delta$, (iii) disorder in the link topology, (iv) initial coordination of the network, and (v) the choice of whether to tune the link tensions (currents) or extensions (pressure drops) (see SI Appendix for results). The values of $\nu$ lie in the range of 0.6-0.8, with the exception of one case of 1.0 for a very large relative change in target response of $\Delta = 1000$ (see Table S1).  We find that the behavior is not well-described by a power law for tuning negative relative changes in target response ($\Delta < 0$) and for tuning small changes in current or tension. The former case is still under investigation, while the latter exception has a simple explanation (see SI Appendix).

Overall, the divergence of the maximum number of tunable targets with system size and the corresponding vanishing of the transition width (indicating the existence of a phase transition) are very robust observations for positive and sufficiently large relative changes in target responses. We note also that both mechanical networks and flow networks exhibit very similar quantitative behavior despite the fact that flow networks are purely topological, requiring no explicit spatial embedding.

The SAT-UNSAT transition of the task complexity problem introduced here represents a new class of discrete constraint-satisfaction transitions due to a new complication that arises in the form of the constraints. When tuning a mechanical network, the removal of links can introduce soft modes, making it impossible to uniquely evaluate the network response, and subsequently tune a given target. Similarly, in a flow network the tuning process can lead to regions being disconnected from the source, making it impossible to tune any target in that region. To avoid such cases, at each step of the tuning process we are forced to exclude specific link removals (see Methods and Materials).
In both mechanical and flow networks, we find that it becomes more and more likely to introduce a soft mode/disconnected region as the task complexity increases. This makes the problem more difficult to tackle both numerically and analytically compared to previously-studied constraint-satisfaction transitions, and may lead to differences in the nature of the transition.

For mechanical functions, a perfectly engineered mechanism (e.g., a pair of chopsticks, which creates a large displacement at the tips in response to strain applied where they are held)
may perform exactly one function superlatively well, but we have shown that more complex network structures are able to adapt to a number of functions that diverges with the system size.
The same argument holds for flow networks: an optimally engineered distribution network is a topological
tree, perfectly suited for a specified task but at the same time ``rigid,'' in the sense that it can not
easily adapt to other tasks. The networks that we have studied are more complex than a pair of chopsticks
or a topological tree, and this allows them to be tuned successfully to perform arbitrarily complex tasks.

Our finding that a disordered network topology allows for tunability may
have relevance to real biological networks.
For example, the development of certain vascular structures in the body of animals is characterized by
the initial appearance of a tightly meshed disordered network of veins (the vascular plexus)
that is subsequently pruned and tuned to its function~\cite[Chapter~1]{Preziosi2003}.
The initial disordered network may be
a prerequisite of the great variability and versatility seen in natural networks. The tuned mechanical networks serve as simple models
for multifunctional allostery in proteins (with a single regulatory
site that can control more than one active site, e.g.,~\cite{Schlessinger1986,Light2013})
or multifunctional metamaterials. Our flow network results
give insight into how to control, for example, blood and
oxygen distribution in vascular systems, or power in an electrical network. Indeed, we find very similar behavior in a flow network with nonplanar topology derived from the UK railroad network. $P_{SAT}$ exhibits a qualitatively similar transition in the number of targets that can be tuned as the networks studied here (see SI Appendix).

Our results raise a number of issues for future investigation. The divergence in the task complexity and vanishing of the transition width with system size are reasonably well-approximated by power laws but may deviate from perfect power laws for larger system sizes (see SI Appendix).  Further work should be carried out to elucidate this behavior. The measured exponents appear to depend on many specific properties of the problems studied. This might be attributable to corrections to scaling or to a more fundamental deviation from power-law scaling.  Also, it is not clear what conditions on the network topology are necessary to observe the transition we see. For example, we expect networks with ring or tree structures to be unable to support very complex tasks. More generally, it has not been investigated how the results depend on network structure/topology and dimensionality nor how they depend on the tuning algorithm.  For instance, the values of $N_T^c /N$ and $\nu$ might be higher for simulated annealing, which explores a wider region of solution space than the minimization algorithm studied here.

One further aspect of our results deserves mention: a simple function that controls only a single pair of target nodes can be achieved in an extremely large number of ways.  We have shown that a task can be complex with $N_T$ randomly chosen target nodes controlled by a single source.  However, if one is only interested in controlling a single target, one can create different paths for its control by choosing any of the $N$ other nodes in the system also to be a target.  Likewise, one could specify a third node to be controlled as well, etc.  That means that there are at least $\sim (N-1)!/(N-N_{T}^{c})! (N_{T}^{c}-1)!$ ways of creating that simple function. Because we find $N_T^c \sim N^\nu$, for $\nu < 1$ this lower bound is smaller than the prediction of $e^{{\cal O}(N)}$ solutions in the large-$N$ limit~\cite{Yan2017}.

Here we studied the limits of the complexity of a single task. It would be interesting to understand how many different tasks can be designed successfully, and whether that is controlled by a similar SAT-UNSAT transition. Finally, we note that for the mechanical and flow networks studied here, the behavior is governed by a discrete Laplacian operator~\cite{Redner2009}--mechanical networks obey force balance on each node and flow networks obey Kirchhoff's law. However, many networks, such as gene regulatory networks, metabolic networks, social networks, etc. are non-conservative. Moreover, the problems we have studied are linear in their couplings but ecological networks or neural networks, for example, are typically nonlinear. It is known that even non-conservative and/or nonlinear networks, such as the Hopfield model and jammed packings, can support SAT-UNSAT transitions as well~\cite{Folli2017, Liu2010}. It would interesting to study systematically how conservation constraints and linearity affect the nature of the transition.

\section*{Materials and Methods}
\subsection*{Linear Response}
Our networks are described by a set of $N$ nodes and $N_E$ edges. The response of a flow network to external stimuli is represented by a pressure $p_i$ on each node $i$. Analogously, the response of a $d$-dimensional mechanical network is the $d$-dimensional displacement vector $\vec{u}_i$ of each node. Each edge linking nodes $i$ and $j$ is characterized by either a conductance or stiffness, denoted $k_{ij}$ in both cases.  For mechanical networks, $k_{ij} = \lambda_{ij} / \ell_{ij}$ where $\lambda_{ij}$ is the stretch modulus per unit length and $\ell_{ij}$ is the rest length. Initially, we set all stretch moduli $\lambda_{ij}$ identically to one. Similarly, for flow networks we set all conductivities $k_{ij}$ to one. Removing an edge $ij$ corresponds to setting $k_{ij}$ to zero, whereas reinserting an edge corresponds to setting $k_{ij}$ back to its original value.

To calculate the response of each type of network, we minimize the corresponding functional. In the case of flow networks, we minimize the power loss through the network,
\begin{align}
P = \sum\limits_{\langle ij\rangle}k_{ij} \qty(p_j -p_i)^2\label{eq:power}
\end{align}
where $\langle ij\rangle$ indicates a sum over all edges. For mechanical networks, we minimize the elastic energy
\begin{align}
E = \frac{1}{2}\sum\limits_{\langle ij\rangle}k_{ij} \qty[\hat{b}_{ij}\cdot(\vec{u}_j -\vec{u}_i)]^2\label{eq:energy}
\end{align}
where $\hat{b}_{ij}$ is a unit vector pointing from node $i$ to node $j$ in the undeformed configuration. The power loss for a flow network can be mapped to the energy of a mechanical network for $d=1$ by mapping the pressure on each node to a one-dimensional displacement~\cite{Tang1987}. In this case, the unit vectors $\hat{b}_{ij}$ are scalars with values of either $\pm 1$, which drop out when squared; the embedding of the network in space does not matter as is be expected for flow networks.

Minimizing~\eqref{eq:power} for a flow network in the presence of externally applied boundary currents $q_i$ on each node $i$, we obtain a system of linear equations characterized by a graph Laplacian $L$,
\begin{align}
L\ket*{p} &= \ket*{q}
\end{align}
where $\ket*{p}$ is an $N$-dimensional vector of node pressures and $\ket*{q}$ is a $N$-dimensional vector of external currents on nodes.  We define the vector $\ket*{i}$ so that the pressure and current on the $i$th node are $p_i = \bra*{i}\ket*{p}$ and $q_i = \bra*{i}\ket*{q}$, respectively. Similarly for mechanical networks, minimizing~\eqref{eq:energy} in the presence of externally applied forces, we obtain
\begin{align}
H\ket*{u} &= \ket*{f}
\end{align}
where $\ket*{u}$ is an $dN$-dimensional vector of node displacements and $\ket*{f}$ is a $dN$-dimensional vector of external forces on nodes. Again we define the $N\times d$ matrix $\ket*{i_d}$ to pick out the displacement and force on the $i$th node, $\vec{u}_i = \bra*{i_d}\ket*{u}$ and $\vec{f}_i = \bra*{i_d}\ket*{f}$. The matrix $H$ is the matrix of second derivatives known as the dynamical or Hessian matrix and can be interpreted as graph Laplacian where each element is a $d\times d$ matrix. We define the $d$-Laplacian, denoted $L_d$, as a generalized version of the standard Laplacian matrix. The case $d=1$ corresponds to the Laplacian of a flow network (or a one-dimensional mechanical network) such that $L_1 = L$, while for $d>1$, $L_d$ is a Hessian for a $d$-dimensional mechanical network, i.e. $L_{d>1} = H$. The $ij$th $d\times d$-block of the $d$-Laplacian is
\begin{align}
\mel*{i_d}{L_d}{j_d} = \left\{\begin{array}{cl}
\sum\limits_{l\neq i}k_{il} \hat{b}_{il}\hat{b}^T_{il} & \qif i=j\\
- k_{ij} \hat{b}_{ij}\hat{b}^T_{ij} &\qif i\neq j
\end{array} \right.
\end{align}
where $k_{ij}$ is nonzero only if edge $ij$ exists.

Consequently, the response of either type of network is calculated by solving the corresponding set of linear equations rewritten as
\begin{align}
L_d\ket*{u} &= \ket*{f}\label{eq:linsys}
\end{align}
where $\ket*{u}$ and $\ket*{f}$ are the appropriate $dN$-dimensional response and source vectors, respectively.
To apply a pressure drop or edge extension source, we use a bordered Laplacian formulation.

\subsection*{Bordered Laplacian Formulation}
Calculating the linear response requires solving~\eqref{eq:linsys}. However, there are two complications. The first is that the Laplacian operator is in general not invertible due to the presence of global degrees of freedom. For a periodic network, in $d$ dimensions, there are $d$ global translational degrees of freedom. Second, we apply edge extension (pressure drop) sources, rather than tension (current) sources. These sources can be applied as constraints on the system. Using a bordered Laplacian formulation, we add a constraint for each global translation and for the source.

First, we define the extension (or pressure drop) of the source as
\begin{align}
e_S &= \hat{b}_S \cdot (\vec{u}_{S_2} - \vec{u}_{S_1}) = \braket*{S}{u}\label{eq:source}
\end{align}
with source nodes $S_1$ and $S_2$. The unit vector $\hat{b}_S$ points from node $S_1$ to $S_2$ and is a scalar in the case of a flow network. The vector $\ket*{S}$ is defined to extract the extension of the source from the full vector of node displacements. We specify the desired extension as $e^*_S$. Additionally, we define the vectors $\ket*{G_i}$ for $i=1,\ldots,d$ corresponding to translations of the entire system uniformly along the $i$th axis. We define the Lagrangian
\begin{align}
\mathcal{L} &= E - \sum\limits_{i=1}^d \lambda_i \braket*{G_i}{u} - \lambda_S (e_S - e^*_S)
\end{align}
where the parameters $\lambda_i$ and $\lambda_S$ are Lagrange multipliers. We include the Lagrange multipliers as additional unknown parameters that must be determined in our calculations. We find solutions by extremizing the Lagrangian with respect to both the displacements and Lagrange multiplier. We rewrite the Lagrangian in matrix form:
\begin{align}
\mathcal{L} &= \frac{1}{2}\mel*{u}{L_d}{u} - \mel*{\lambda_G}{G^T}{u} - \lambda_S (\braket*{S}{u} - e^*_S).
\end{align}
The vector $\ket*{\lambda_G}$ is size $d$ with elements $\braket*{i}{\lambda_G} = \lambda_i$ and $G$ is a size $dN\times d$ matrix with columns $G\ket*{i} = \ket*{G_i}$.
In this context we can further condense notation, writing the Lagrangian as
\begin{align}
\mathcal{L} &= \frac{1}{2}\mel*{\overline{u}}{\overline{L}_d}{\overline{u}}
\end{align}
where we define the bordered Laplacian $\overline{L}_d$ as a block matrix of second derivatives of the Lagrangian.
\begin{align}
\overline{L}_d &=
\qty(\begin{array}{ccc}
L_d & -G & \ket*{S}\\
-G^T & 0 & 0\\
\bra*{S} & 0 & 0
\end{array}).
\end{align}
We also define the bordered displacement and force vectors $\ket*{\overline{u}}$ and $\ket*{\overline{f}}$, respectively, each of size $dN+d+1$ as
\begin{align}
\ket*{\overline{u}} = \qty(\begin{array}{c}
\ket*{u} \\
\ket*{\lambda_G}\\
\lambda_S
\end{array}), \quad \ket*{\overline{f}} = \qty(\begin{array}{c}
\ket*{f} \\
0\\
-e_S^*.
\end{array}).
\end{align}
As a result, the system of equations we must solve is now $\overline{L}_d\ket*{\overline{u}} = \ket*{\overline{f}}$.
%\begin{align}
%\overline{L}_d\ket*{\overline{u}} = \ket*{\overline{f}}.
%\end{align}
The bordered Laplacian is invertible due to the presence of the constraints and solving this equation is straightforward.

\subsection*{Tuning Loss Function}

Framed according to \eqref{eq:constraint}, the problem of tuning a complex task can be viewed as a constraint-satisfaction problem. The goal is to find a set of stiffnesses (conductivities) that simultaneously satisfy each constraint in \eqref{eq:constraint}. To study this problem numerically, we recast it as an optimization problem in the style of Ref.~\cite{Franz2017}, in which we define an objective function that penalizes deviation of the system's behavior from the desired multifunctionality. Thus, we introduce the loss function
\begin{align}
\mathcal{F}[\{k_{ij}\}] = \frac{1}{2} \sum_{\alpha=1}^{N_T} r_\alpha^2 \Theta(-r_{\alpha}), \label{eq:cost}
\end{align}
which is a function of the set of all the spring constants (conductivities) $\{k_{ij}\}$, and is composed of a sum over the set of $N_T$ target edges to be tuned. For each target edge $\alpha$ we define the residual
\begin{align}
r_{\alpha} = \frac{\eta_{\alpha} - \eta^{(0)}_\alpha}{\eta^{(0)}_\alpha } - \Delta  . \label{eq:residual}
\end{align}
which measures how close each target is to being tuned successfully. The Heaviside function $\Theta(-r_\alpha)$ is included so that if $r_{\alpha} > 0$ , i.e., the response ratio has increased at least by the desired proportion $\Delta$, then the residual does not contribute to the loss function.

\subsection*{Optimization Method}
Our method for tuning a network involves minimizing the loss function in~\eqref{eq:cost}.  In the spirit of~\cite{Goodrich2015,Rocks2017}, our optimization consists of removing or reinserting previously removed edges from the network one at a time, modifying the network topology in discrete steps. More specifically, we use a greedy algorithm in which we remove or reinsert the edge which minimizes the loss function at each step. This requires a calculation of the new response for each possible move. 

Suppose we have a network whose stiffnesses at the current step are $\{k_{ij}\}$ for all valid $ij$ where some $k_{ij}$ might already be zero, having been removed at previous steps. Our goal is to measure the change in response when the stiffness of edge $ij$ is changed by an amount $\Delta k_{ij}$.  We note that the Laplacian can be decomposed as
\begin{align}
L_d = Q K Q^T
\end{align}
where the equilibrium (or incidence) matrix $Q$ of size $dN \times N_E$ defines the mapping of nodes to edges\cite{Pellegrino1993, Redner2009} and $K$ is a size $N_E \times N_E$ diagonal matrix of edge stiffnesses such that $\mel*{ij}{K}{lm} = k_{ij}\delta_{ij, lm}$. We can define a bordered incidence matrix $\overline{Q}$ by appending $d+1$ rows of zeros to $Q$, giving us a corresponding decomposition of the bordered Laplacian $\overline{L}_d = \overline{Q} K  \overline{Q}^T$. The change in response is
\begin{align}
\ket*{\Delta\overline{u}} &= \qty[\qty(\overline{L}_d+\Delta \overline{L}_d)^{-1} - \overline{L}_d^{-1}]\ket*{\overline{f}}
\end{align}
with the corresponding change in the bordered Laplacian $\Delta \overline{L}_d = \Delta k_{ij}\ketbra*{q_{ij}}{q_{ij}}$
%\begin{align}
%\Delta \overline{L}_d &= \Delta k_{ij}\ketbra*{q_{ij}}{q_{ij}}
%\end{align}
with the vector $\ket*{q_{ij}} = \overline{Q}\ket*{ij}$.
We now need to calculate the inverse of the updated bordered Laplacian. This can be done using the Sherman-Morrison formula~\cite{Sherman1950}
\begin{align}
\qty(\overline{L}_d+\Delta \overline{L}_d)^{-1} &= \overline{L}_d^{-1} - \frac{\overline{L}_d^{-1}\Delta k_{ij}\ketbra*{q_{ij}}{q_{ij}}\overline{L}_d^{-1}}{1 - \Delta k_{ij} \mel{q_{ij}}{\overline{L}_d^{-1}}{q_{ij}}}
\end{align}
The change in response is then
\begin{align}
\ket*{\Delta\overline{u}} &= - \frac{\overline{L}_d^{-1}\Delta k_{ij}\ketbra*{q_{ij}}{q_{ij}}\overline{L}_d^{-1}\ket*{\overline{f}}}{1 - \Delta k_{ij} \mel{q_{ij}}{\overline{L}_d^{-1}}{q_{ij}}}
\end{align}
The new response is then used to calculate an updated loss function.

In order to reduce numerical error and maintain the numerical invertibility of the bordered Laplacian, we define the quantity
\begin{align}
S_{ij}^2 \equiv 1 - \Delta k_{ij} \mel{q_{ij}}{\overline{L}_d^{-1}}{q_{ij}}
\end{align}
If $S_{ij}^2$ is less than $10^{-4}$, we do not remove an edge. This quantity can be shown to be the contribution of an edge to the states-of-self-stress in mechanical systems \cite{Sussman2016, Hexner2018}. By ensuring that every removed edge has some contribution to the states-of-self-stress, then by Maxwell-Calladine counting, we are guaranteed that no zero modes are introduced \cite{Calladine1978}.

We repeatedly add or remove edges until  either the loss function is explicitly zero (i.e., all constraints are satisfied), or the relative change in the objective function is less than $10^{-8}$. 

\section*{Acknowledgements}
We thank S. Franz for instructive discussions, along with C. P. Goodrich, D. Hexner and N. Pashine. This research was supported by the US Department of Energy, Office of Basic Energy Sciences, Division of Materials Sciences and Engineering under Awards DE-FG02-03ER46088 (S.R.N.) and DE-FG02-05ER46199 (J.W.R.), the University of Chicago MRSEC NSF DMR-1420709 (S.R.N.), the Simons Foundation for the collaboration Cracking the Glass Problem via awards 454945 (A.J.L.) and 348125 (S.R.N.), Simons Investigator Award 327939 (A.J.L.), the National Science Foundation under Award No. PHY-1554887 (E.K. and H.R.), the Burroughs Welcome Career Award (E.K. and H.R.), and by a National Science Foundation Graduate Fellowship (J.W.R).

\bibliographystyle{pnas-new}
\bibliography{limits_multifunc_preprint}

\renewcommand{\theequation}{S\arabic{equation}}
\setcounter{equation}{0}

\renewcommand\thefigure{S\arabic{figure}}
\setcounter{figure}{0}

\renewcommand\thetable{S\arabic{table}}
\setcounter{table}{0}

\section*{Supporting Information (SI)}

\subsection*{Variations of the network tuning problem}
We performed many variations of the standard network tuning problem presented in the main text. The default simulation parameters we used were a pressure (flow networks) or extension (mechanical networks) source, a target relative change in response of $\Delta = 0.1$, and an average node coordination of $Z = 2N_E /N \approx 5.0$. For both flow and mechanical networks, we studied the two cases in which the two source nodes were connected by a single edge and where the two source nodes were chosen randomly from all the nodes, giving 4 cases altogether that we discuss in the main text. Table~\ref{tab:sims} shows the many variations on these parameters that we explored, along with the resulting power law exponents where applicable and the corresponding figures showing the satisfiability transitions and scaling of the transition position $N_T^c$ and width $w$. Each set of simulations had at least 128 simulations per data point to calculate the satisfaction probability. The first section of the table shows the data for the four cases discussed in the main text. The second and third sections show configurations for flow and mechanical networks, respectively for higher values of the target relative change $\Delta$. The fourth and fifth sections show results for random networks with $Z \approx 4.1$ (close to isostaticity for mechanical networks) and initially perfect triangular lattices. The sixth section shows configurations for tuning the response with global strains applied to mechanical networks. The seventh section shows results for tuning a target current in response to a current source (flow networks) or target tension in response to a tension source (mechanical networks). Finally, the last two sections show results for a negative relative change in the target response for flow and mechanical networks. All variations on the tuning problem show qualitatively similar power law behavior except for tuning small changes in current or tension (see Section ``Tuning target current'') or negative desired relative changes in target response, $\Delta < 0.0$) (see Section ``Tuning negative target change $\Delta$''). 

\subsubsection*{Tuning target current}
In Table~\ref{tab:sims},  we do not list exponents for flow networks tuned for target current nor mechanical networks tuned for target tension with $\Delta = 0.1$. As seen in Figs.~\ref{fig:current}(A) and (B), these cases do not result in the typical power law behavior seen elsewhere. Instead we find that it is almost always possible to achieve the desired response. This stems from the fact that the current in flow networks or tension in mechanical networks can be trivially increased in magnitude by simply removing the source edge. Typically, the source edge acts as either a resistor or a spring in parallel to the rest of the network, diverting a significant fraction of all current or tension through that edge. If the source edge is removed, then the magnitude of the current or tension is increased without changing the sign. We find that this increase is always enough to satisfy at least a 10\% change in magnitude ($\Delta = 0.1$), but not enough to satisfy $\Delta = 1.0$ in flow networks nor $\Delta = 10.0$ in mechanical networks. For these latter cases, the resulting transitions revert back to the typical behavior seen elsewhere.

\subsubsection*{Tuning negative target change $\Delta$}
The last two sections of Table~\ref{tab:sims} contain the sets of variables we tested for the alternate case of a negative relative target response, $\Delta < 0$. The resulting transitions are depicted in Figs.~\ref{fig:flow_neg_delta} and \ref{fig:mech_neg_delta}. For these cases, we flip the inequality in Eq.~(1), resulting in the constraints
\begin{align}
\frac{\eta_\alpha - \eta_\alpha^{(0)}}{\eta_\alpha^{(0)}} \leq \Delta, \alpha = 1,\ldots,N_T.
\end{align}
Note that $\Delta >0$ corresponds to increasing the magnitude of the response without changing the sign, $-1 < \Delta < 0$ corresponds to decreasing the magnitude of the response without changing the sign, and $\Delta < -1$ corresponds to tuning target responses of the opposite sign from the source. For $\Delta<0$, we do not always see a simple power law behavior for reasons that are still under investigation.

%{\color{red}
\subsubsection*{Transition in the UK rail network}\label{sec:UK_rail}
In order to demonstrate that real transportation networks possess properties similar to the jammed packing topologies considered in
the main text, we analyze a part of the UK Rail network presented in
Ref.~\cite{Gallotti2015}.
While the rail network is well-connected, it contains several parts which
are connected through bridges, edges whose removal creates two
disconnected components. Because a source located in one such component
does not influence the other, we remove all bridges in the rail network graph
and focus on the remaining largest connected component, which
consists of $N=2030$ nodes and $N_E = 3868$ edges. Both for node pair and edge sources,
the probability $P_{SAT}$ generally
exhibits qualitatively similar behavior to the jammed packing networks considered in the
main text (see Fig.~\ref{fig:UK_rail}). There is a slight difference in the functional form of the drop in $P_{SAT}$ for the case of an edge source. This difference arises because the transportation network is significantly less interconnected than networks derived from
jammed packings. As a result, certain edges exhibit
zero pressure drop for a given source. If such a target edge
is randomly chosen, it is automatically set to be satisfied.
This case occurs particularly often for edge sources.
%}

\subsection*{Transition power law fitting and deviations}

Fig.~\ref{fig:deviation} demonstrates the deviations from power laws for the four systems displayed in Fig.~3. We have plotted the fractional difference of each measured point, $N_T^c$ or $w$, from its fitted power law function $f(N)$ and $g(N)$, respectively, as a function of system size $N$.  In both cases our fitted function is of the form $A N^\alpha$ where $A$ and $\alpha$ are our fit parameters. Both the data sets for $N_T^c$ and $w$ are fit simultaneously with the same power $\alpha$, but different coefficients $A$, resulting in a total of three fit parameters. Error bars have been estimated by dividing the uncertainty in $N_T^c$ or $w$ by the respective fit function at that point. It is apparent that the simple power law form does not perfectly match the underlying data.

\subsection*{Satisfaction probability error bars}
Each data point of the various satisfaction probability plots is representative of a binomial distribution 
\begin{align}
p_i \sim \mathrm{Binomial}(n_i, \hat{p}_i)
\end{align}
where $n_i$ is the number samples and $\hat{p}_i$ is the fraction of successful tuning attempts. To calculate the error bars depicted in the various satisfaction probability plots, we use the Wilson score interval~\cite{Wilson1927}
\begin{align}
p_i^\pm = \frac{1}{1+\frac{1}{n_i}z^2}\qty[\hat{p}_i+\frac{1}{2n_i}z^2\pm z\sqrt{\frac{1}{n_i}\hat{p}_i(1-\hat{p}_i)+\frac{1}{4n_i^2}z^2}]
\end{align}
with a z-score of $z=1$. This gives us an estimate of the uncertainty for each data point which is analogous to the standard deviation for a Gaussian distribution. However, since the probability is restricted between zero and one, the error bars are not necessarily symmetric.

\subsection*{Satisfaction probability curve fitting}

The satisfaction probability curves depicted in Fig.~2, along with many of the supplemental figures, were estimated using smoothing splines constructed from a basis of cubic B-splines. The procedure for constructing the splines and estimating the smoothing parameter were drawn with some modification from Ref.~\citep[Chapter~9.2]{Wahba1990}.

To generate an estimate of a satisfaction probability curve, we start with a set of $n$ satisfaction probabilities $y_i$ each generated for a corresponding number of targets $x_i$ where $i$ goes from 1 to $n$. Each satisfaction probability counts the fraction of successfully tuned networks from a collection of $n_i$ samples. Our goal is to find a function $p(x)$ which approximates the underlying function sampled by the data. Since we do not know what functional form we should use, we would like to approximate this function using a spline. However, the function $p(x)$ should be limited to the interval $[0,1]$, while splines are not typically limited in this way. Therefore, we write $p(x)$ in terms of a more general function as
\begin{align}
p(x) &= \frac{e^{S(x)}}{1 + e^{S(x)}}
\end{align}
where $S(x)$ is the spline function which can take on any real value.

\subsubsection*{B-spline approximation}

In terms of B-splines, the approximating spline function $S(x)$ is written
\begin{align}
S(x) &= \sum\limits_{i=1}^m c_iB_i^k(x)
\end{align}
with $m$ coefficients $c_i$ and degree-$k$ basis splines $B_i^k(x)$. The coefficients are the fit parameters we would like to estimate. 

We must address the specific choices made in the use of B-splines. First, we choose to use cubic splines ($k=3$). One knot is chosen for each data point plus an extra $k$ at the lowest and highest values of $x$ for padding. This gives us a total of $m = n+2k$ knots, 
\begin{align}
t_i &= \left\{\begin{array}{cl}
x_1 & \qif   0 < i \leq k\\
x_{i-k} & \qif  k < i \leq n+k\\
x_n & \qif n+k < i < n+2k
\end{array}\right.
\end{align}
The result is $m = n+2k - (k+1)$ basis splines with corresponding coefficients.

\subsubsection*{B-spline coefficient estimation}

Typically, one would employ a least squares approach to calculate the spline coefficients. However, this assumes that each data point is drawn from some normal distribution, while we know in this case they are drawn from a set of binomial distributions
\begin{align}
y_i \sim \mathrm{Binomial}(n_i, p(x_i))
\end{align}
Carrying out a standard maximum likelihood estimation, the corresponding log-likelihood of the binomially distributed data is
\begin{align}
\mathcal{L}(y) &= \frac{1}{n}\sum\limits_{i=1}^n n_i\qty[y_i\log p(x_i) + (1-y_i) \log(1-p(x_i))]
\end{align}
In terms of $S(x)$, the log-likelihood can be written
\begin{align}
\mathcal{L}(y) &= \frac{1}{n}\sum\limits_{i=1}^n\qty[b(x_i) - y_iS(x_i)]
\end{align}
up to a constant with
\begin{align}
b(x_i) &= n_i \log\qty(1+e^{S(x_i)})
\end{align}
To implement smoothing, we introduce a term with penalty parameter $\lambda$ which penalizes the square of the curvature of $S(x)$. This gives us the penalized generalized linear model
\begin{align}
I_\lambda(c) &=  \frac{1}{n}\sum\limits_{i=0}^{n-1}\qty[n_i\log\qty(1+e^{S(x_i)}) - y_iS(x_i)] + \lambda \int\limits_{x_0}^{x_{n-1}} dx [S''(x)]^2\label{eq:smooth}
\end{align}

\subsubsection*{Smoothing parameter}

Next we must choose a good value for $\lambda$. This is accomplished using a generalized cross-validation (GCV) approach, allowing us to choose $\lambda$ in an agnostic manner. Using GCV effectively chooses $\lambda$ so that the approximating spline curve changes as little as possible if an arbitrary subset of data is left out of the fit. For the sake of convenience, we write
\begin{align}
S(x) &= \braket*{c}{B^k(x)}
\end{align}
where $c_i = \braket*{i}{c}$ and $B^k_i(x) = \braket*{i}{B^k(x)}$ are vectors of size $m$. We also write
\begin{align}
\Sigma_{ij} &= \int\limits_{x_0}^{x_{n-1}} dx  B^k_i(x)B^k_j(x)
\end{align}
Finally, we minimize the generalized cross-validation function
\begin{align}
V(\lambda)_{GCV} &= \frac{\sum\limits_{i=1}^n\qty[D^{-\frac{1}{2}}_i(y_i-\mu_i)]^2}{\frac{1}{n}\tr^2(I-A)}\label{eq:GCV}
\end{align}
with
\begin{align}
\mu_i &= b'(S(x_i))\\
D_i &= b''(S(x_i))
\end{align}
and
\begin{align}
A_{ij} = D^{\frac{1}{2}}_i \bra*{B^k(x_i)} \qty[\sum\limits_{l=1}^n  D_l \ketbra*{B^k(x_l)}{B^k(x_l)} + 2\lambda \Sigma]^{-1}\ket*{B^k(x_j)} D^{\frac{1}{2}}_j
\end{align}
The size $n\times n$ matrix  $I$ is the identity. When testing a particular value of $\lambda$, the values $c_i$ are always chosen to minimize \eqref{eq:smooth} for that $\lambda$. Therefore, the spline coefficients are treated as a function of $\lambda$. 

When minimizing \eqref{eq:GCV}, there may sometimes be extraneous minima at $\lambda = 0$ or $\lambda=\infty$. Since we would like some degree of smoothing, we never choose the minimum at zero. Also, moderate smoothing is generally preferable to infinite smoothing, so if a local minimum exists for finite $\lambda$, it is chosen even if it is not the global minimum. 

\subsection*{Transition measurements}

We use the spline approximations of each satisfaction probability curve in order to estimate the positions and widths of each satisfiability transition. The center of the transition is simply chosen as the number of targets $N_T^c$ such that the probability of success is exactly $50\%$,  $P_{SAT}(N_T^c) = 0.5$. The width of the transition $w$ is found by first finding the number of targets corresponding to success rates of $25\%$ and $75\%$ and taking their differences, $w = P_{SAT}^{-1}(0.75) - P_{SAT}^{-1}(0.25)$. In order to weight each point correctly when finding the power law scaling of the transition properties, we utilized Monte Carlo resampling to estimate uncertainty~\cite{press2007numerical}. To find the uncertainty values for $N_T^c$ and $w$ for a particular curve, each data point for that curve is resampled from its underlying binomial distribution. The spline approximation is then recalculated for this new set of data points and new values of $N_T^c$ and $w$ are extracted. This process is repeated numerous times, resulting in a distribution of value of $N_T^c$ and $w$. The uncertainty is then calculated by finding the standard deviation of of these distributions.

\begin{figure}%[tbhp!]
\centering
\includegraphics[width=\textwidth]{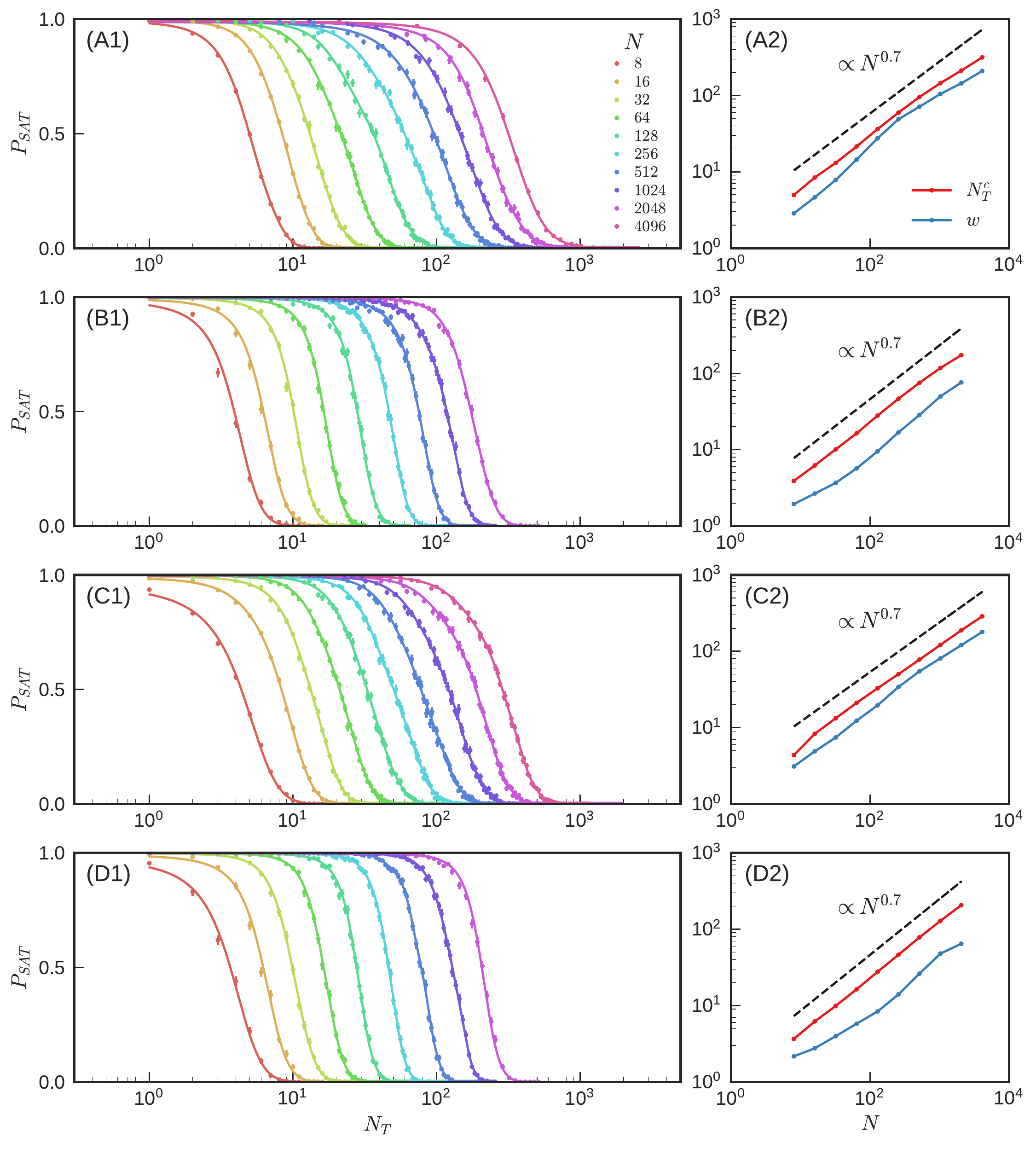}
\caption{Satisfaction probability and scaling of the transition position and width for the four main cases shown in the main text: (A) flow networks and (B) mechanical networks with an edge source and (C) flow networks and (D) mechanical networks with a node pair source. See Table~\ref{tab:sims} for more details.}
\label{fig:default}
\end{figure}

\begin{figure}%[tbhp!]
\centering
\includegraphics[width=\textwidth]{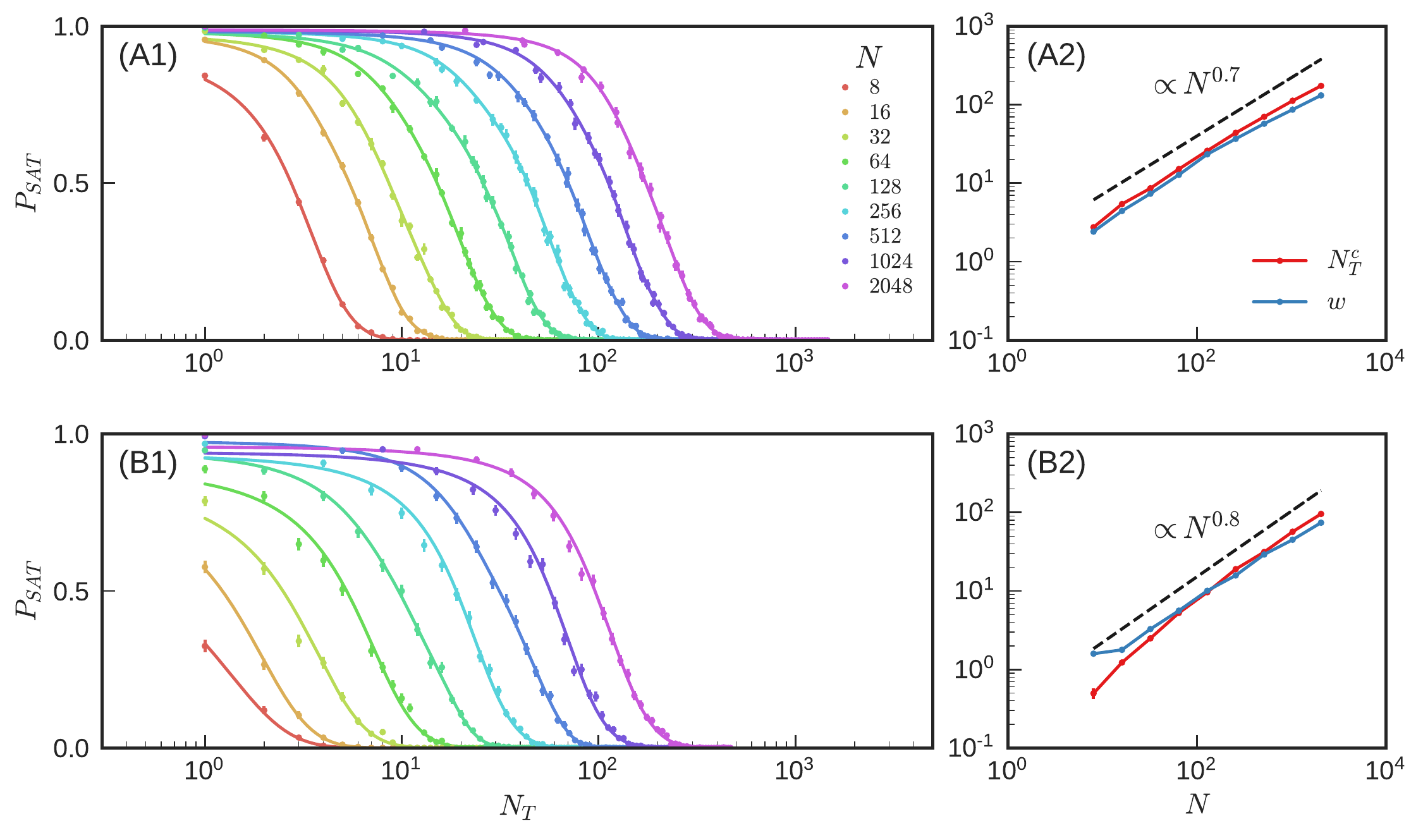}
\caption{Satisfaction probability and scaling of the transition position and width for flow networks with desired relative change in target response of (A) $\Delta=1.0$ and (B) $\Delta=10.0$. See Table~\ref{tab:sims} for more details.}
\label{fig:flow_delta}
\end{figure}

\begin{figure}%[tbhp!]
\centering
\includegraphics[width=\textwidth]{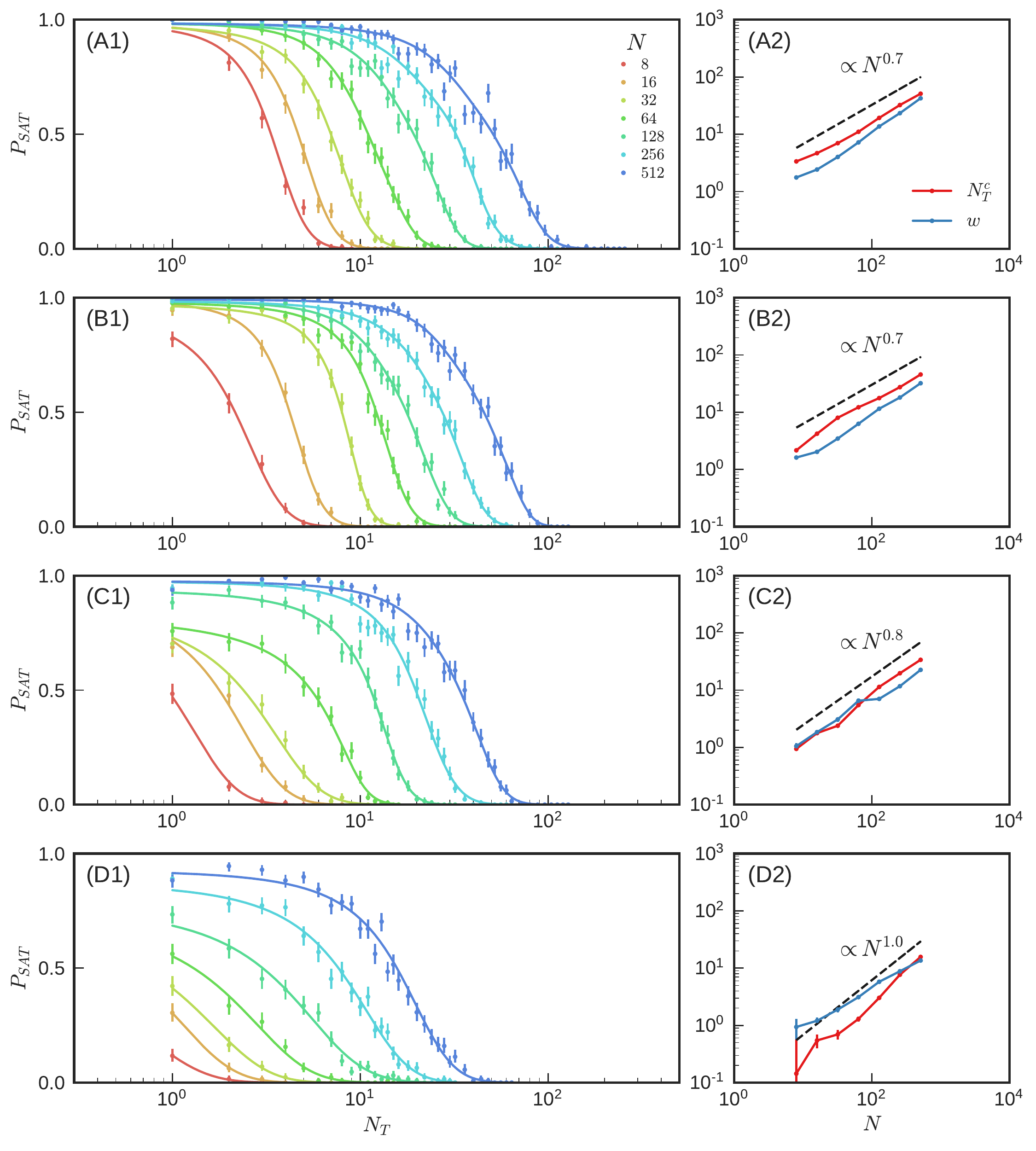}
\caption{Satisfaction probability and scaling of the transition position and width for mechanical networks with desired relative change in target response of (A) $\Delta=1.0$, (B) $\Delta=10.0$, (C) $\Delta=100.0$, and (D) $\Delta=1000.0$. See Table~\ref{tab:sims} for more details.}
\label{fig:mech_delta}
\end{figure}

\begin{figure}%[tbhp!]
\centering
\includegraphics[width=\textwidth]{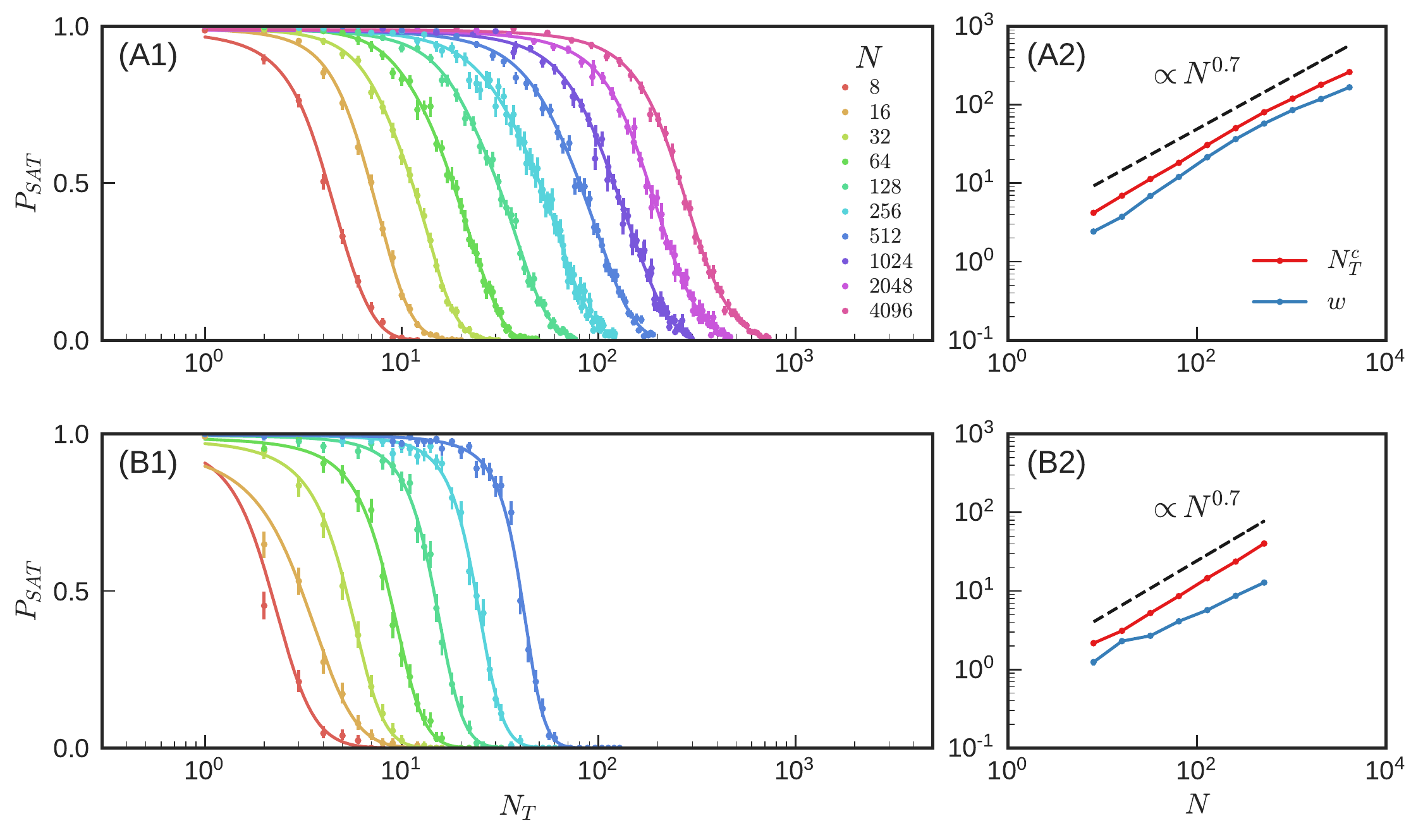}
\caption{Satisfaction probability and scaling of the transition position and width for (A) flow networks and (B) mechanical networks with average connectivity of $Z \approx 4.1$, lower than the default of $Z \approx 5.0$. See Table~\ref{tab:sims} for more details.}
\label{fig:low_DZ}
\end{figure}

\begin{figure}%[tbhp!]
\centering
\includegraphics[width=\textwidth]{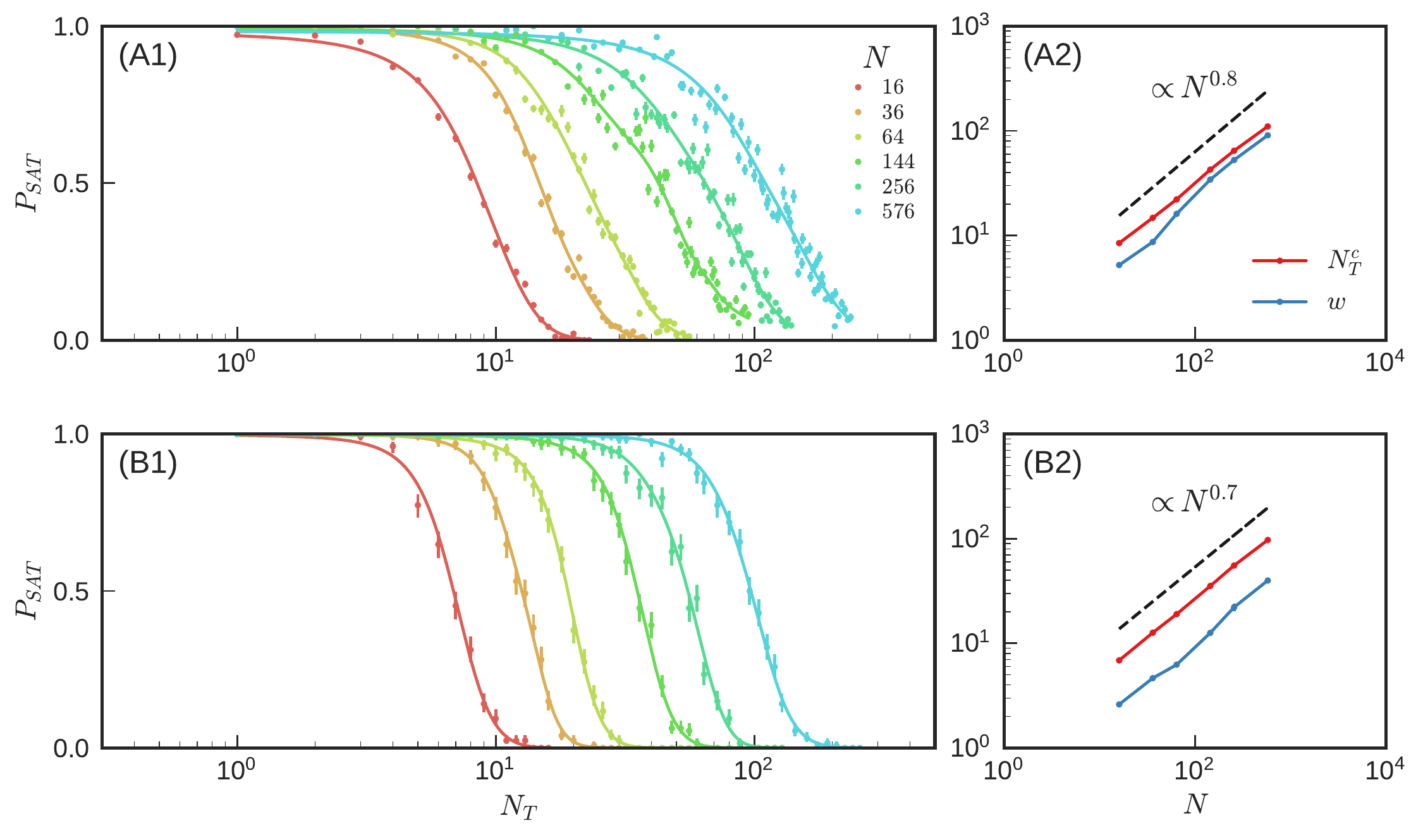}
\caption{Satisfaction probability and scaling of the transition position and width for (A) flow networks and (B) mechanical networks on an ordered triangular lattice. See Table~\ref{tab:sims} for more details.}
\label{fig:tri}
\end{figure}

\begin{figure}%[tbhp!]
\centering
\includegraphics[width=\textwidth]{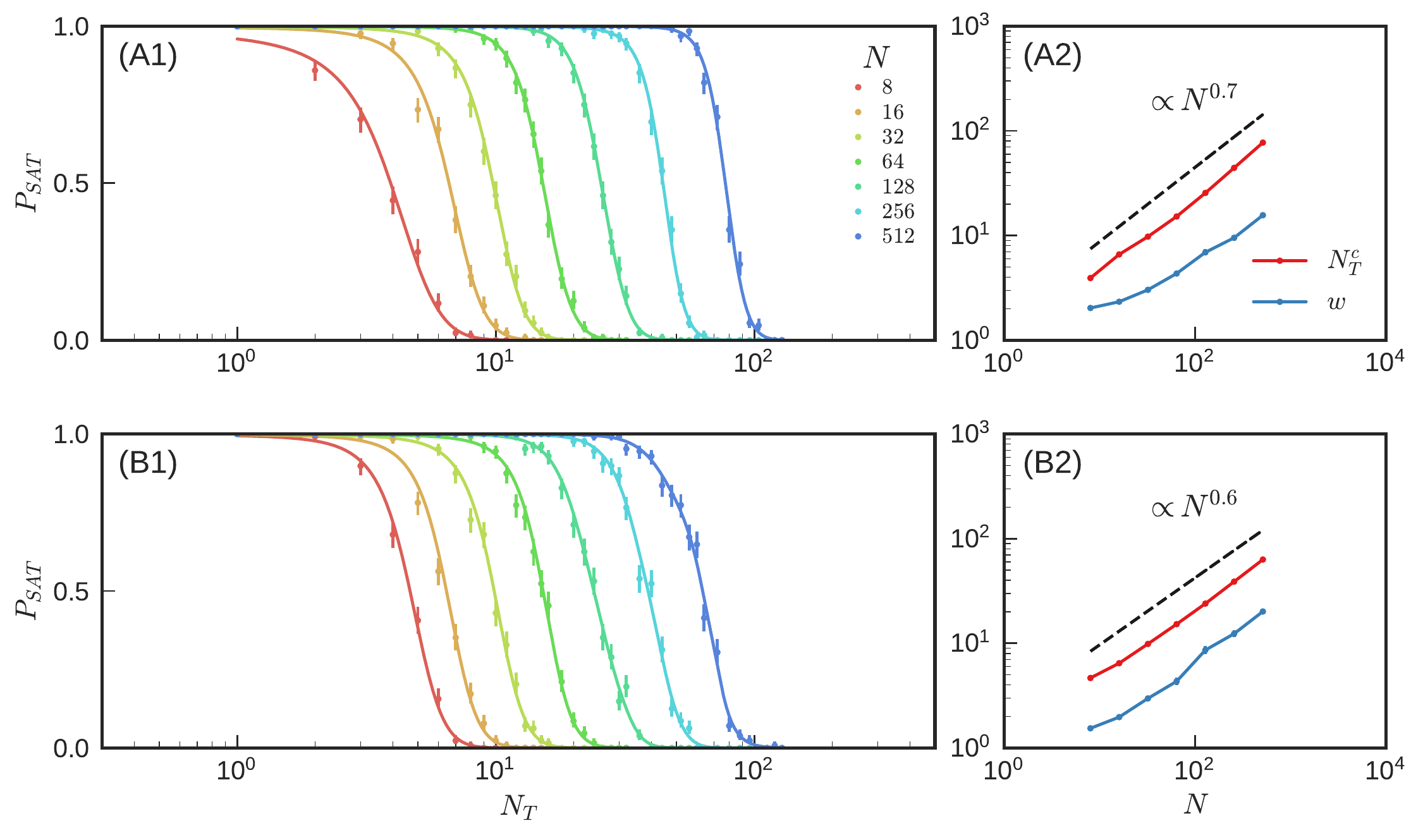}
\caption{Satisfaction probability and scaling of the transition position and width for mechanical networks with global (A) shear and (B) compression sources. See Table~\ref{tab:sims} for more details.}
\label{fig:global}
\end{figure}

\begin{figure}%[tbhp!]
\centering
\includegraphics[width=\textwidth]{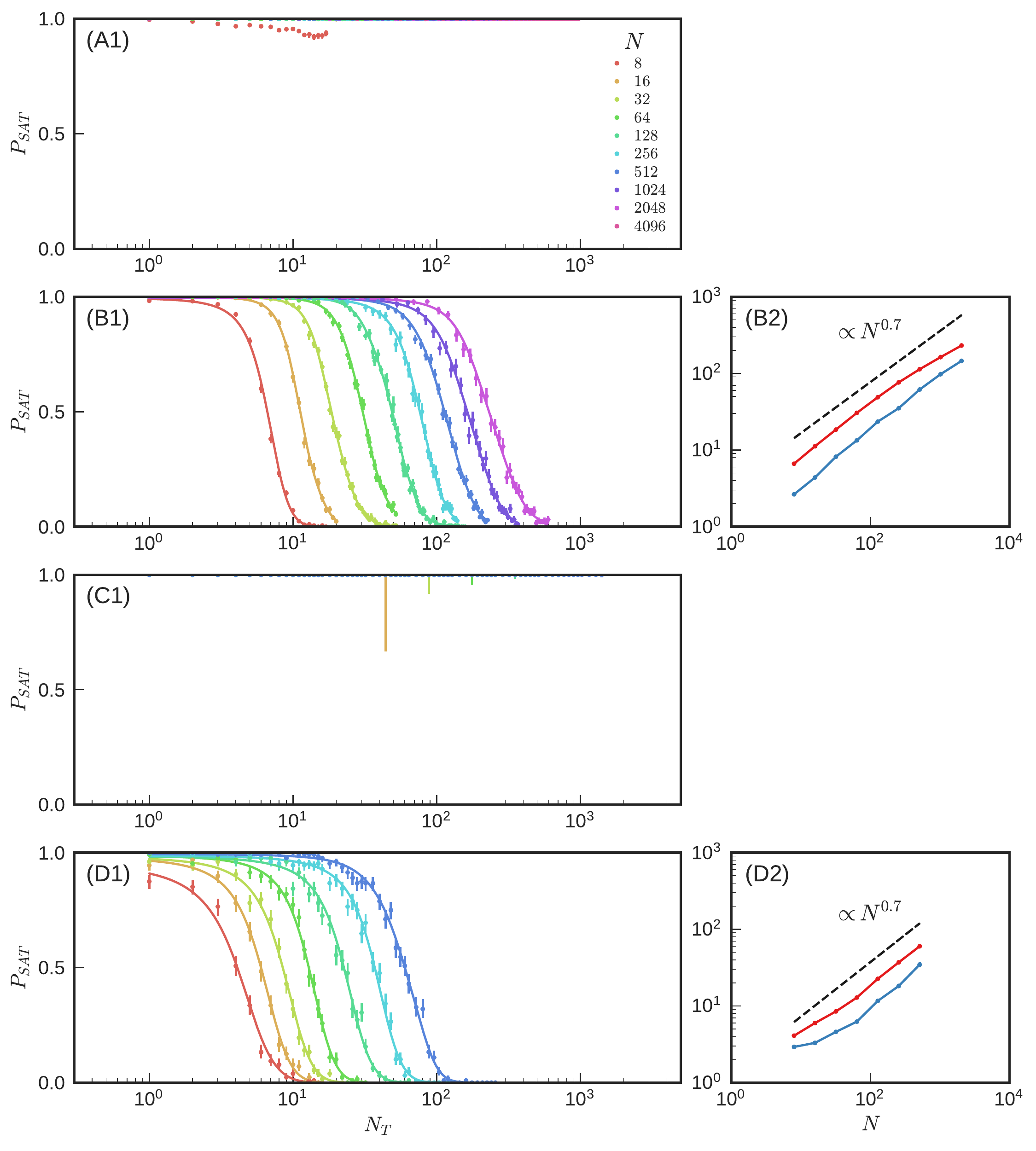}
\caption{Satisfaction probability and scaling of the transition position and width for flow networks tuned for target current with (A) $\Delta = 0.1$ and $\Delta = 1.0$ and mechanical networks tuned for target tension with (C) $\Delta = 0.1$ and (D) $\Delta = 10.0$. See Table~\ref{tab:sims} for more details. Large error bars reflect a lack of available networks with enough edges to measure $P_{SAT}$ for large $N_T$. }
\label{fig:current}
\end{figure}

\begin{figure}%[tbhp!]
\centering
\includegraphics[width=\textwidth]{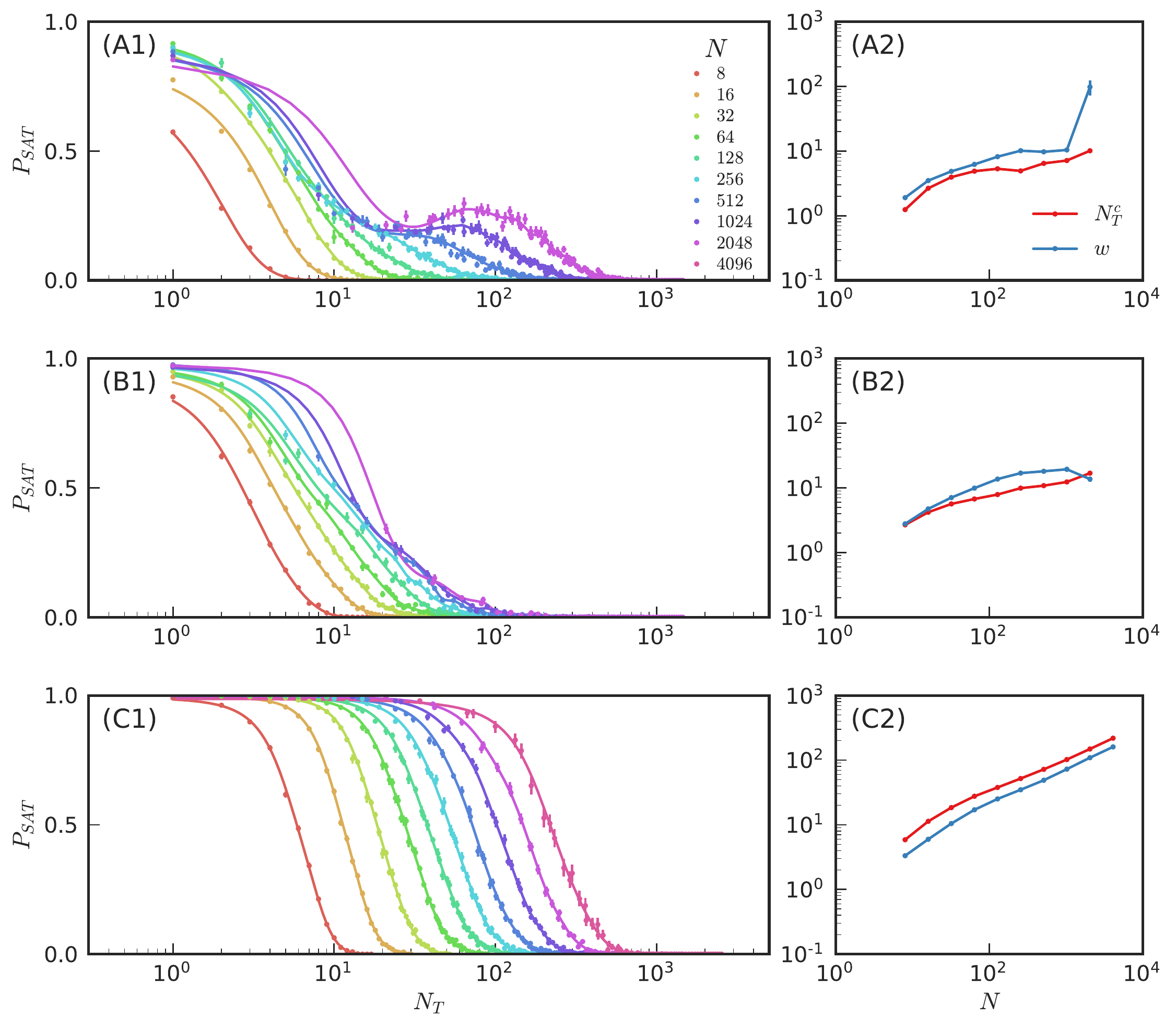}
\caption{Satisfaction probability and scaling of the transition position and width for flow networks tuned for a negative relative change in target response of (A) $\Delta=-0.5$, (B) $\Delta=-1.0$, and (C) $\Delta=-1.5$. See Table~\ref{tab:sims} for more details.}
\label{fig:flow_neg_delta}
\end{figure}

\begin{figure}%[tbhp!]
\centering
\includegraphics[width=\textwidth]{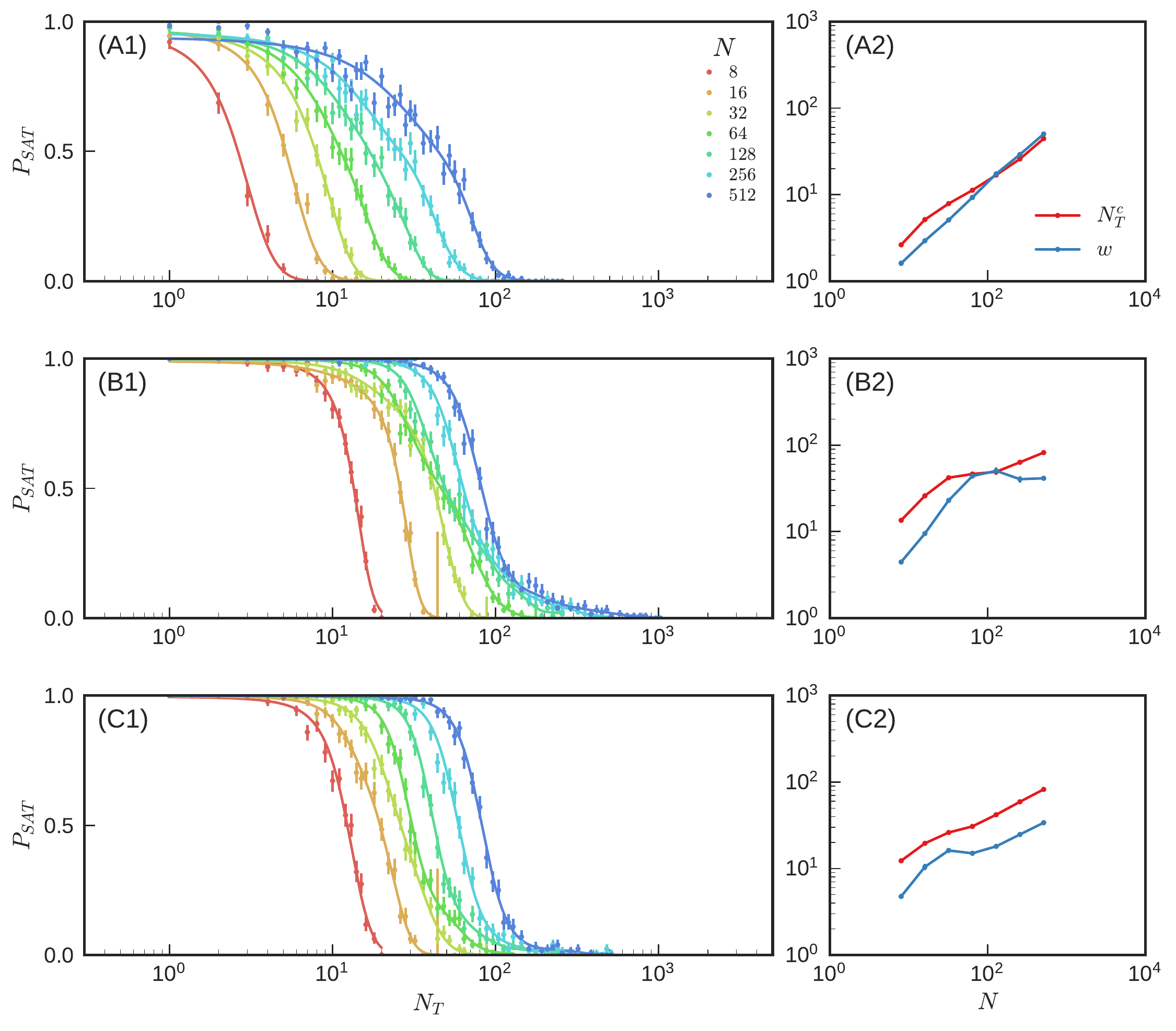}
\caption{Satisfaction probability and scaling of the transition position and width for mechanical networks tuned for a negative relative change in target response of (A) $\Delta=-0.5$, (B) $\Delta=-1.0$, and (C) $\Delta=-1.5$. See Table~\ref{tab:sims} for more details. Large error bars reflect a lack of available networks with enough edges to measure $P_{SAT}$ for large $N_T$. }
\label{fig:mech_neg_delta}
\end{figure}

\begin{figure}%[tbhp!]
\centering
\includegraphics[width=0.7\textwidth]{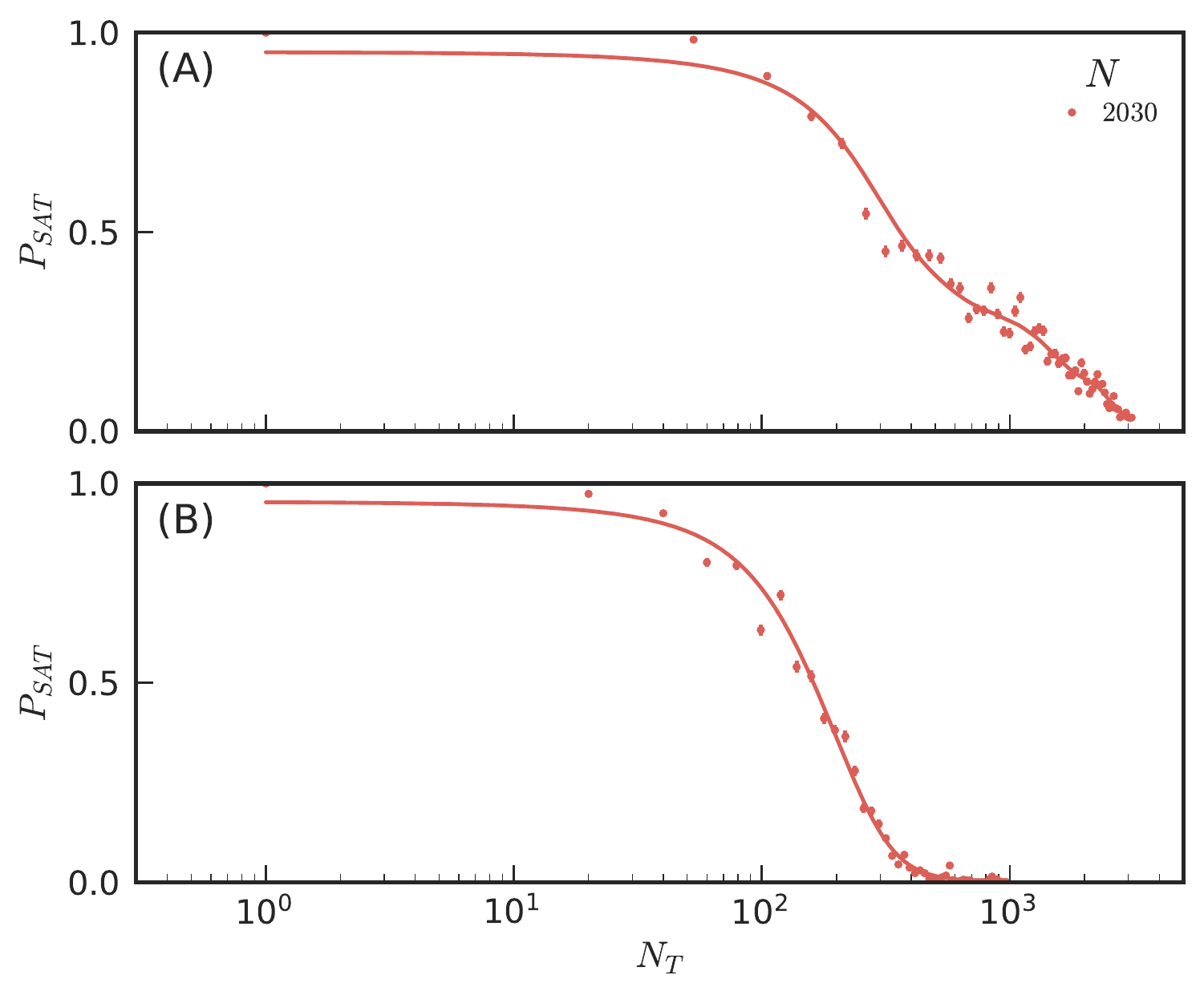}
\caption{Satisfaction probability of flow networks derived from real UK railroad networks with (A) an edge source and (B) a node pair source. See Section~\ref{sec:UK_rail} for more details.}
\label{fig:UK_rail}
\end{figure}

\begin{figure}%[tbhp!]
\centering
\includegraphics[width=0.7\textwidth]{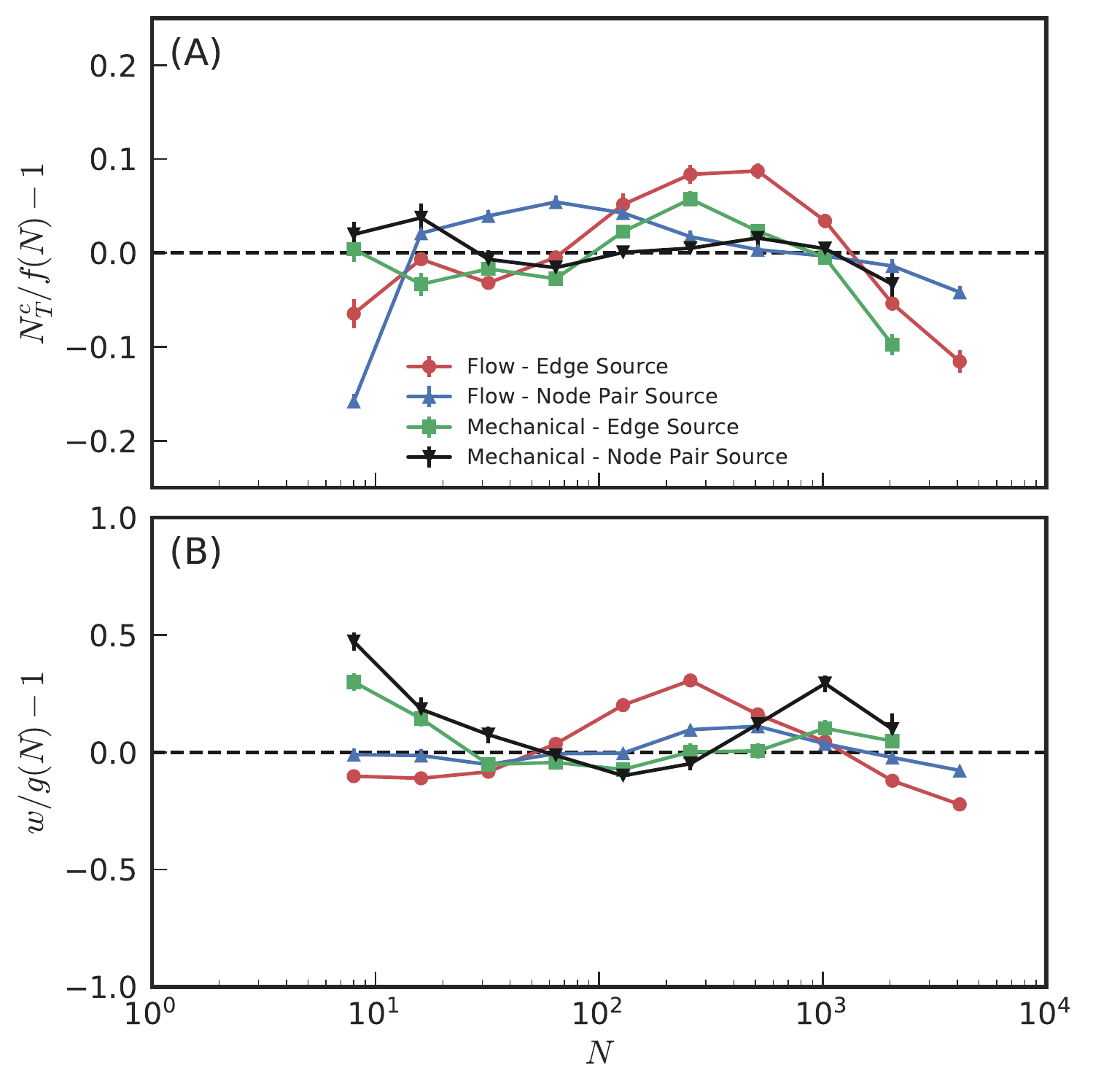}
\caption{Deviations of the power laws in Fig.~3 of the main text from power laws of the form $f(N) = AN^\alpha$ for the transition position and $g(N) = BN^\alpha$ for the transition width. Error bars for the position and width have been rescaled by dividing by $f(N)$ or $g(N)$, respectively.}
\label{fig:deviation}
\end{figure}

\begin{table}
\setlength{\tabcolsep}{12pt}
\centering
\caption{Variations of tuning problem and corresponding transition exponents}
\resizebox{\columnwidth}{!}{
\begin{tabular}{lllrlrl}
\thead{Physical\\ System} & \thead{Source\\ Properties} & \thead{Target\\ Properties} & \thead{Target\\ Change $\Delta$} & \thead{Network\\ Properties} & \thead{Transition\\ Position}  & Figure(s)\\
\midrule
Flow & Edge Pressure Drop & Edge Pressure Drop & $0.1$ & Random - $Z \approx 5.0$ & $0.7$ & 1(A), 2(A),3, \ref{fig:default}(A)\\
Mechanical & Edge Extension & Edge Extension & $0.1$ & Random - $Z \approx 5.0$ & $0.7$ & 1(B), 2(B), 3, \ref{fig:default}(B)\\
Flow & Node Pair Pressure Drop & Edge Pressure Drop & $0.1$ & Random - $Z \approx 5.0$ & $0.6$ & 1(C)\, 3, \ref{fig:default}(C)\\
Mechanical & Node Pair Extension & Edge Extension & $0.1$ & Random - $Z \approx 5.0$ &  $0.7$ & 1(D)\, 3, \ref{fig:default}(D)\\
\hline
Flow & Edge Pressure Drop & Edge Pressure Drop & $\bf{1.0}$ & Random - $Z \approx 5.0$ & $0.7$  & \ref{fig:flow_delta}(A)\\
Flow & Edge Pressure Drop & Edge Pressure Drop & $\bf{10.0}$ & Random - $Z \approx 5.0$ & $0.8$  & \ref{fig:flow_delta}(A)\\
\hline
Mechanical & Edge Extension & Edge Extension & $\bf{1.0}$ & Random - $Z \approx 5.0$ & $0.7$ & \ref{fig:mech_delta}(A)\\
Mechanical & Edge Extension & Edge Extension & $\bf{10.0}$ & Random - $Z \approx 5.0$ & $0.7$ & \ref{fig:mech_delta}(B)\\
Mechanical & Edge Extension & Edge Extension & $\bf{100.0}$ & Random - $Z \approx 5.0$ & $0.8$ & \ref{fig:mech_delta}(C)\\
Mechanical & Edge Extension & Edge Extension & $\bf{1000.0}$ & Random - $Z \approx 5.0$ & $1.0$ & \ref{fig:mech_delta}(D)\\
\hline
Flow & Edge Pressure Drop & Edge Pressure Drop & $0.1$ & \textbf{Random -} $\bf{Z \approx 4.1}$ &  $0.7$  & \ref{fig:low_DZ}(A)\\
Mechanical & Edge Extension & Edge Extension & $0.1$ & \textbf{Random -} $\bf{Z \approx 4.1}$ &  $0.7$ & \ref{fig:low_DZ}(B)\\
\hline
Flow & Edge Pressure Drop & Edge Pressure Drop & $0.1$ & \textbf{Triangular Lattice} &  $0.8$ & \ref{fig:tri}(C)\\
Mechanical & Edge Extension & Edge Extension & $0.1$ & \textbf{Triangular Lattice} & $0.7$ & \ref{fig:tri}(D)\\
\hline
Mechanical & \textbf{Global Shear} & Edge Extension & $0.1$ & Random - $Z \approx 5.0$ &  $0.7$   & \ref{fig:global}(A)\\
Mechanical & \textbf{Global Expansion} & Edge Extension & $0.1$ & Random - $Z \approx 5.0$ & $0.6$  & \ref{fig:global}(B)\\
\hline
Flow & Edge Current & \textbf{Edge Current} & $0.1$ & Random - $Z \approx 5.0$ & N/A & \ref{fig:current}(A)\\
Flow & Edge Current & \textbf{Edge Current} & $\bf{1.0}$ & Random - $Z \approx 5.0$ & $0.7$  & \ref{fig:current}(B)\\
Mechanical & Edge Tension & \textbf{Edge Tension} & $0.1$ & Random - $Z \approx 5.0$ & N/A & \ref{fig:current}(C)\\
Mechanical & Edge Tension & \textbf{Edge Tension} & $\bf{10.0}$ & Random - $Z \approx 5.0$ & $0.7$ & \ref{fig:current}(D)\\
\hline
Flow & Edge Pressure Drop & Edge Pressure Drop & $\bf{-1.5}$ & Random - $Z \approx 5.0$ & N/A & \ref{fig:flow_neg_delta}(A)\\
Flow & Edge Pressure Drop & Edge Pressure Drop & $\bf{-1.0}$ & Random - $Z \approx 5.0$ & N/A & \ref{fig:flow_neg_delta}(B)\\
Flow & Edge Pressure Drop & Edge Pressure Drop & $\bf{-0.5}$ & Random - $Z \approx 5.0$  & N/A & \ref{fig:flow_neg_delta}(C)\\
\hline
Mechanical & Edge Extension & Edge Extension & $\bf{-1.5}$ & Random - $Z \approx 5.0$ & N/A  & \ref{fig:mech_neg_delta}(A)\\
Mechanical & Edge Extension & Edge Extension & $\bf{-1.0}$ & Random - $Z \approx 5.0$ & N/A & \ref{fig:mech_neg_delta}(B)\\
Mechanical & Edge Extension & Edge Extension & $\bf{-0.5}$ & Random - $Z \approx 5.0$ & N/A & \ref{fig:mech_neg_delta}(C)\\
\bottomrule
\end{tabular}
}
\newline\newline

\raggedright
N/A indicates power law estimates not applicable due to lack of transition, or clearly non-power-law-like behavior.
Bold text indicates changes from default parameters.\label{tab:sims}
\end{table}

\end{document}